\documentclass[aps,prl,reprint]{revtex4-2}
\usepackage{graphics}     
\usepackage{graphicx}      
\usepackage{longtable}     
\usepackage{url}          
\usepackage{bm}          
\usepackage{mathtools,amsmath,amssymb,lipsum,xcolor}

\usepackage[utf8]{inputenc}
\usepackage[]{natbib}

\DeclarePairedDelimiter\bra{\langle}{\rvert}
\DeclarePairedDelimiter\ket{\lvert}{\rangle}
\DeclarePairedDelimiterX\braket[2]{\langle}{\rangle}{#1 \delimsize\vert #2}

\begin{document}
\title{The fractional Landau-Lifshitz-Gilbert equation}
\author         {R.C. Verstraten$^{1}$, T. Ludwig$^{1}$, R.A. Duine$^{1,2}$, C. Morais Smith$^{1}$}
\affiliation    {$^{1}$Institute for Theoretical Physics, Utrecht University, Princetonplein 5, 3584CC Utrecht, The Netherlands \\
$^{2}$Department of Applied Physics, Eindhoven University of Technology,
P.O. Box 513, 5600 MB Eindhoven, The Netherlands}
\date{\today}

\begin{abstract}
The dynamics of a magnetic moment or spin are of high interest to applications in technology. Dissipation in these systems is therefore of importance for improvement of efficiency of devices, such as the ones proposed in spintronics. A large spin in a magnetic field is widely assumed to be described by the Landau-Lifshitz-Gilbert (LLG) equation, which includes a phenomenological Gilbert damping. Here, we couple a large spin to a bath and derive a generic (non-)Ohmic damping term for the low-frequency range using a Caldeira-Leggett model. This leads to a fractional LLG equation, where the first-order derivative Gilbert damping is replaced by a fractional derivative of order $s\ge 0$. We show that the parameter $s$ can be determined from a ferromagnetic resonance experiment, where the resonance frequency and linewidth no longer scale linearly with the effective field strength.
\end{abstract}



\maketitle

\textit{Introduction.}--- The magnetization dynamics of materials has attracted much interest because of its technological applications in spintronics, such as data storage or signal transfer~\cite{mayergoyz2009nonlinear,harder2016electrical,barman2020magnetization}. The right-hand rule of magnetic forces implies that the basic motion of a magnetic moment or macrospin $\bm{S}$ in a magnetic field $\bm{B}$ is periodic precession. However, coupling to its surrounding (e.g., electrons, phonons, magnons, and impurities) will lead to dissipation, which will align $\bm{S}$ with $\bm{B}$. 

Spintronics-based devices use spin waves to carry signals between components~\cite{Guo2021}. Contrary to electronics, which use the flow of electrons, the electrons (or holes) in spintronics remain stationary and their spin degrees of freedom are used for transport. This provides a significant advantage in efficiency, since the resistance of moving particles is potentially much larger than the dissipation of energy through spins. The spin waves consist of spins precessing around a magnetic field and they are commonly described by the Landau-Lifshitz-Gilbert (LLG) equation~\cite{lakshmanan2011fascinating}. This phenomenological description also includes Gilbert damping, which is a term that slowly realigns the spins with the magnetic field. Much effort is being done to improve the control of spins for practical applications~\cite{awschalom2007challenges}. Since efficiency is one of the main motivations to research spintronics, it is important to understand exactly what is the dissipation mechanism of these spins. 

Although the LLG equation was first introduced phenomenologically, since then it has also been derived from microscopic quantum models~\cite{koopmans2005unifying,duine2007functional}. Quantum dissipation is a topic of long debate, since normal Hamiltonians will always have conservation of energy. It can be described, for instance, with a Caldeira-Leggett type model~\cite{Caldeira1981,Caldeira1983physa,Caldeira1983annphys,Caldeira2012,Weiss2012}, where the Hamiltonian of the system is coupled to a bath of harmonic oscillators. These describe not only bosons, but any degree of freedom of an environment in equilibrium. These oscillators can be integrated out, leading to an effective action of the system that is non-local and accounts for dissipation. The statistics of the bath is captured by the spectral function $J(\omega)$, which determines the type of dissipation.
For a linear spectral function (Ohmic bath), the first-order derivative Gilbert damping is retrieved.

The spectral function is usually very difficult to calculate or measure, so it is often assumed for simplicity that the bath is Ohmic. However, $J(\omega)$ can have any continuous shape. Hence, a high frequency cutoff is commonly put in place, which sometimes justifies a linear expansion. However, a general expansion is that of an $s$ order power-law, where $s$ could be any positive real number. A spectral function with such a power-law is called non-Ohmic, and we refer to $s$ as the ``Ohmicness'' of the bath. 
It is known that non-Ohmic baths exist~\cite{Anders2022,groeblacher2015observation,Abdi2018analog,Kehrein1996spinboson,Wilner2015subohmic,Nalbach2010ultraslow,Paavola2009environment,Wu2013dynamics,Jeske2018effects,Lemmer2018trapped} and that they can lead to equations of motion that include fractional derivatives~\cite{lutz2012fractional,Metzler2000,Mainardi1997,de2014review,Verstraten2021}. Because fractional derivatives are non-local, these systems show non-Markovian dynamics which can be useful to various applications~\cite{hilfer2000applications,dalir2010applications,gardiner2004quantum}.

Here, we show that a macroscopic spin in contact with a non-Ohmic environment leads to a fractional LLG equation, where the first derivative Gilbert damping gets replaced by a fractional Liouville derivative. Then, we explain how experiments can use ferromagnetic resonance (FMR) to determine the Ohmicness of their environment from resonance frequency and/or linewidth. This will allow experiments to stop using the Ohmic assumption, and use equations based on measured quantities instead. The same FMR measurements can also be done with anisotropic systems. Aligning anisotropy with the magnetic field may even aid the realization of measurements, as this can help reach the required effective field strengths. In practice, the determination of the type of environment is challenging, since one needs to measure the coupling strength with everything around the spins. However, with the experiment proposed here, one can essentially measure the environment through the spin itself. Therefore, the tools that measure spins can now also be used to determine the environment. This information about the dissipation may lead to improved efficiency, stability, and control of applications in technology.



\textit{Derivation of a generalized LLG equation.}--- We consider a small ferromagnet that is exposed to an external magnetic field. Our goal is to derive an effective equation of motion for the magnetization. For simplicity, we model the magnetization as one large spin (macrospin) $\hat{\bm{S}}$. Its Hamiltonian (note that we set $\hbar$ and $k_B$ to one) reads $\hat H_s =  \bm{B} \cdot \hat{\bm{S}} - K \hat{S}_z^2$, where the first term (Zeeman) describes the coupling to the external magnetic field $\bm{B}$, and the second term accounts for (axial) anisotropy of the magnet. However, since a magnet consists of more than just a magnetization, the macrospin will be in contact with some environment. Following the idea of the Caldeira-Leggett approach \cite{Caldeira1981,Caldeira1983physa,Caldeira1983annphys,Caldeira2012,caldeira1985influence,Weiss2012}, we model the environment as a bath of harmonic oscillators, $\hat H_b = \sum_\alpha \hat{\bm{p}}_\alpha^2/2m_\alpha + m_\alpha \omega_\alpha^2 \hat{\bm{x}}_\alpha^2/2$, where $\hat{\bm{x}}_\alpha$ and $\hat{\bm{p}}_\alpha$ are position and momentum operators of the $\alpha$-th bath oscillator with mass $m_\alpha$ and eigenfrequency $\omega_\alpha>0$. Furthermore, we assume the coupling between the macrospin and the bath modes to be linear, $\hat H_c =  \sum_\alpha \gamma_\alpha\, \hat{\bm{S}} \cdot \hat{\bm{x}}_\alpha$, where $\gamma_\alpha$ is the coupling strength between macrospin and the $\alpha$-th oscillator. Thus, the full Hamiltonian of macrospin and environment is given by $\hat H = \hat H_s + \hat H_c + \hat H_b$.

Next, we use the Keldysh formalism in its path-integral version~\cite{Kamenev2011, altlandsimons2010}, which allows us to derive an effective action and, by variation, an effective quasi-classical equation of motion for the macrospin. For the path-integral representation of the macrospin, we use spin coherent states~\cite{altlandsimons2010} $\ket{g} = \exp(- i\phi \hat S_z) \exp (-i \theta \hat S_y) \exp(- i\psi \hat S_z)\, \ket{\uparrow}$, where $\phi$, $\theta$, and $\psi$ are Euler angles and $\ket{\uparrow}$ is the eigenstate of $\hat S_z$ with the maximal eigenvalue $S$. Spin coherent states provide an intuitive way to think about the macrospin as a simple vector $\bm{S} = \bra{g}\hat{\bm{S}}\ket{g} = S\, (\sin \theta \cos \phi, \sin \theta \sin \phi, \cos \theta)$ with constant length $S$ and the usual angles for spherical coordinates $\theta$ and $\phi$. For spins, the third Euler angle $\psi$ presents a gauge freedom, which we fix as in Ref. \cite{Shnirman2015} for the same reasons explained there.

After integrating out the bath degrees of freedom, see Sup.~Mat.~\cite{sm} for details, we obtain the Keldysh partition function $\mathcal{Z} = \int Dg\, \exp[i \mathcal S]$, with the Keldysh action
\begin{align}
	\mathcal S = &\oint\! dt\, \big[ S\, \dot \phi\, (1 - \cos \theta) - \bm{B}_\mathrm{eff}(S_z) \cdot \bm{S}  \big] \nonumber \\
	&- \oint\! dt \oint\! dt'\ \bm{S}(t)\, \alpha(t-t')\, \bm{S}(t')\ . \label{eq:krots}
\end{align}
The first term, called Berry connection, takes the role of a kinetic energy for the macrospin; it arises from the time derivative acting on the spin coherent states $(-i \partial_t\bra{g} ) \ket{g} = S\, \dot \phi\, (1- \cos \theta)$. The second term is the potential energy of the macrospin, where we introduced an effective magnetic field, $\bm{B}_\mathrm{eff}(S_z)  = \bm{B} - K S_z\, \bm{e}_z $, given by the external magnetic field and the anisotropy.  The third term arises from integrating out the bath and accounts for the effect of the environment onto the macrospin; that is, the kernel function $\alpha (t-t')$ contains information about dissipation and  fluctuations. Dissipation is described by the retarded and advanced components $\alpha^{R/A}(\omega)=\sum_\alpha (\gamma_\alpha^2/2m_\alpha \omega_\alpha^2)\, \omega^2/[ (\omega \pm i0)^2 -\omega_\alpha^2]$, whereas the effect of fluctuations is included in the Keldysh component, $\alpha^K(\omega) = \coth( \omega/2 T)\, [\alpha^R(\omega) - \alpha^A(\omega)]$. This is determined by the fluctuation-dissipation theorem, as we assume the bath to be in a high-temperature equilibrium state~\cite{Kubo1966, Kamenev2011, altlandsimons2010}.

From the Keldysh action, Eq. \eqref{eq:krots}, we can now derive an equation of motion for the macrospin by taking a variation. More precisely, we can derive quasi-classical equations of motion for the classical components of the angles $\theta$ and $\phi$ by taking the variation with respect to their quantum components~\cite{footnote1}. The resulting equations of motion can be recast into a vector form and lead to a generalized LLG equation
\begin{equation}
	\dot{\bm{S}}(t)= \bm{S}(t) \times \left[-\bm{B}_\mathrm{eff}[S_z(t)] +  \int_{-\infty}^{t}\!\!\!\! dt'\,  \alpha(t-t')\, \bm{S} (t') +  \bm{\xi}(t) \right], \label{eq:gllg}
\end{equation}
with the dissipation kernel~\cite{footnote2} given by 
\begin{equation}
	\alpha (\omega)=  \int_{-\infty}^\infty \frac{d\varepsilon}{\pi} \frac{ \varepsilon  J(\varepsilon) }{(\omega + i0)^2 -\varepsilon^2}\ , \label{eq:alphatilde}
\end{equation}
where we introduced the bath spectral density $J(\omega)= \sum_\alpha (\pi \gamma_\alpha^2/2m_\alpha \omega_\alpha)\,  \delta(\omega-\omega_\alpha)$~\cite{Kamenev2011,sm}. The last term in Eq.~\eqref{eq:gllg} contains a stochastic field $\bm{\xi}(t)$, which describes fluctuations (noise) caused by the coupling to the bath; the noise correlator for the components of $\bm{\xi}(t)$ is given by $\langle \xi_m(t)  \xi_n(t') \rangle = -2 i\, \delta_{mn}\, \alpha^K(t-t')$. Next, to get a better understanding of the generalized LLG equation, we consider some examples of bath spectral densities.



\textit{Fractional Landau-Lifshitz-Gilbert equation.}--- For the generalized LLG equation \eqref{eq:gllg}, it is natural to ask: In which case do we recover the standard LLG equation? We can recover it for a specific choice of the bath spectral density $J(\omega)$, which we introduced in Eq.~\eqref{eq:alphatilde}. Roughly speaking, $J(\omega)$ describes two things: first, in the delta function $\delta(\omega- \omega_\alpha)$, it describes at which energies $\omega_\alpha$ the macrospin can interact with the bath; second, in the prefactor $\pi \gamma_\alpha^2/2m_\alpha \omega_\alpha$, it describes how strongly the macrospin can exchange energy with the bath at the frequency $\omega_\alpha$. In our simple model, the bath spectral density is a sum over $\delta$-peaks because we assumed excitations of the bath oscillators to have an infinite life time. However, also the bath oscillators will have some dissipation of their own, such that the $\delta$-peaks will be broadened. If, furthermore, the positions of the bath-oscillator frequencies $\omega_\alpha$ is dense on the scale of their peak broadening, the bath spectral density becomes a continuous function instead of a collection of $\delta$-peaks. In the following, we focus on cases where the bath spectral density is continuous.

Since the bath only has positive frequencies, we have $J(\omega\le0)=0$. Even though $J(\omega)$ can have any positive continuous shape, one might assume that it is an approximately linear function at low frequencies; that is, 
\begin{equation}
	J(\omega) = \alpha_1\, \omega\, \Theta(\omega)\Theta(\Omega_c-\omega) , 
\end{equation}
where $\Theta(\omega) = 1$ for $\omega >0$ and $\Theta(\omega) = 0$ for $\omega <0$ and $\Omega_c$ is some large cutoff frequency of the bath such that we have $\omega_\text{system}\ll T\ll \Omega_c$. Reservoirs with such a linear spectral density are also known as Ohmic baths. Inserting the Ohmic bath spectral density back into Eq.~\eqref{eq:alphatilde}, while sending $\Omega_c\to\infty$, we recover the standard LLG equation,
\begin{equation}
	\dot{\bm{S}}(t) = \bm{S}(t) \times \left[  -\bm{B}_\mathrm{eff}[S_z(t)] +\alpha_1  \dot{\bm{S}}(t) +  \bm{\xi}(t)\right], \label{eq:llg}
\end{equation}
where the first term describes the macrospin's precession around the effective magnetic field, the second term---known as Gilbert damping---describes the  dissipation of the macrospin's energy and angular momentum into the environment, and the third term describes the fluctuations with $\langle \xi_m(t) \xi_n(t') \rangle = 4\alpha_1  T\, \delta_{mn}\, \delta(t-t')$, which are related to the Gilbert damping by the fluctuation-dissipation theorem. Note that the same results can be obtained without a cutoff frequency by introducing a counter term, which effectively only changes the zero-energy level of the bath, see Sup. Mat.~\cite{sm} for details.

The assumption of an Ohmic bath can sometimes be justified, but is often chosen out of convenience, as it is usually the simplest bath type to consider. To our knowledge, there has been little to no experimental verification whether the typical baths of magnetizations in ferromagnets are Ohmic or not. To distinguish between Ohmic and non-Ohmic baths, we need to know how the magnetization dynamics depends on that difference. Hence, instead of the previous assumption of a linear bath spectral density (Ohmic bath), we now assume that the bath spectral density has a power-law behavior at low frequencies,
\begin{equation}
	J(\omega) = \tilde{\alpha}_s \omega^s\, \Theta(\omega)\Theta(\Omega_c-\omega)\ , \label{eq:spectral}
\end{equation} 
where we refer to $s$ as Ohmicness parameter~\cite{footnote3}. It is convenient to define  $\alpha_s= \tilde{\alpha}_s/ \sin (\pi s/2)$ and we should note that the dimension of $\alpha_s$ depends on $s$. For $s=1$ we recover the Ohmic bath. Correspondingly, baths with $s<1$ are called sub-Ohmic and baths with $s>1$ are called super-Ohmic. For $0<s<2$ and $\Omega_c\to \infty$, we find the fractional LLG equation
\begin{equation}
	\dot{\bm{S}}(t) = \bm{S} \times \left[ -\bm{B}_\mathrm{eff}[S_z(t)] +\alpha_s  D_t^s \bm{S}(t) +  \bm{\xi}(t)\right], \label{eq:fllg}
\end{equation}
where $D_t^s$ is a (Liouville) fractional time derivative of order $s$ and the noise correlation is given by $\langle \xi_m(t) \xi_n(t') \rangle =  2\alpha_s  T\, \delta_{mn}\, (t-t')^{-s}/\Gamma(1-s)$; for a detailed calculation, see the Sup.~Mat.~\cite{sm}. Indeed, in the limit of $s \rightarrow 1$, we recover the regular LLG equation.
The fractional LLG equation~\eqref{eq:fllg} seems quite similar to the standard LLG equation~\eqref{eq:llg}. However, the first-order time derivative in the Gilbert damping is replaced by a fractional $s$-order time derivative in the fractional Gilbert damping; this has drastic consequences for the dissipative macrospin dynamics.



\begin{figure}[b]
	\center
	\includegraphics[width=\columnwidth]{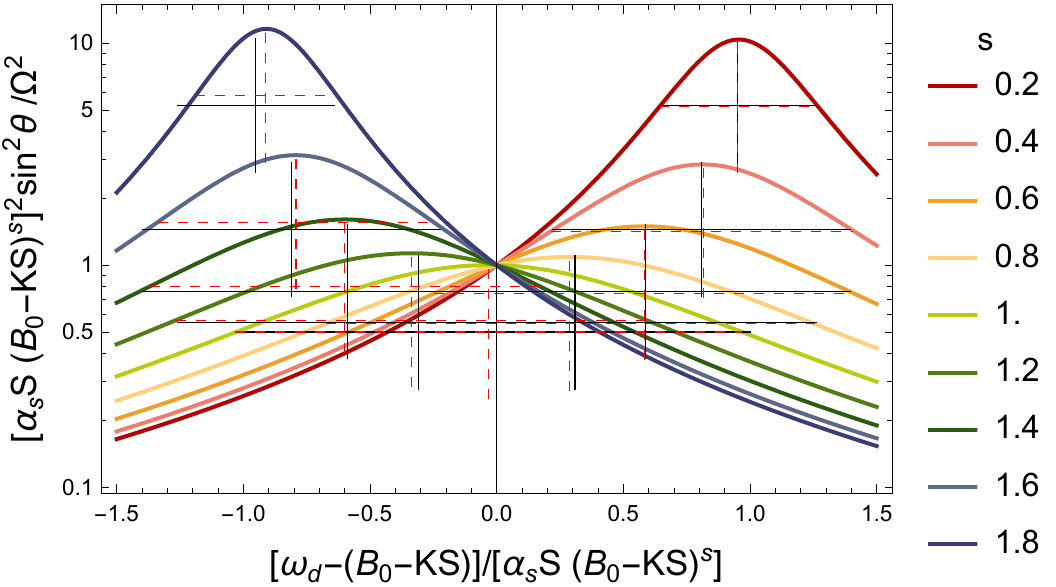}
	\caption{
		A lin-log plot of the amplitude $\sin^2 \theta$ as a function of driving frequency $\omega_d$ plotted in dimensionless units for several values of $s$. The resonance peaks change, depending on $s$. The resonance frequency $\omega_\text{res}$ and linewidth $\Delta_{H/2}$ have been overlayed with crosses. The red dashed crosses have been calculated numerically, whereas the black solid crosses are the derived results from Eqs. \eqref{eq:kres} to \eqref{eq:fwhm}.}
\label{fig:fmr1}
\end{figure}

\textit{Fractional Gilbert damping.}--- Fractional derivatives have a long history~\cite{hilfer2000applications}, and many different definitions exist for varying applications~\cite{hilfer2000applications,Mainardi1997,de2014review}. From our microscopic model, we found the Liouville derivative~\cite{footnote4}, which is defined as
\begin{equation}
	D_t^s \bm{S}(t) = \frac{d^n}{dt^n} \frac{1}{\Gamma(n-s)} \int_{-\infty}^t dt' (t-t')^{n-1-s} \bm{S}(t'),
\end{equation}
where $n$ is the integer such that $n\le s<n+1$. This can be interpreted as doing a fractional integral followed by an integer derivative. The fractional integral is a direct generalization from the rewriting of a repeated integral, by reversing the order of integration into a single one, which leads to extra powers of $(t-t')$.  

To provide some intuition to the effects of fractional friction, we propose a thought experiment. Suppose an object is traveling at constant speed, then $x(t)=v t$. Hence, the friction force acting on the object goes like $ D_t^s x(t)\propto  v t^{1-s}$. Therefore, we find three regimes. For $s=1$, the friction is constant in time. For $s<1$, the friction force increases with time. Hence, longer movements will be less common. For $s>1$, the friction decreases with time, so longer movements will be more likely once set in motion. 

Within the fractional LLG equation, we thus see two important new regimes. For $s<1$ (sub-Ohmic), the friction is more likely to relax (localize) the spin (e.g. sub-diffusion) towards the $\bm{B}$-field direction. For small movements, the friction could be very small, whereas it would greatly increase for bigger movements. This could describe a low dissipation stable configuration. For $s>1$ (super-Ohmic), the friction could reduce as the spin moves further, which in other systems is known to cause L\'evy-flights or super-diffusion~\cite{Metzler2000,lutz2012fractional,dubkov2008levy}. This might lead the system to be less stable, but can potentially also greatly reduce the amount of dissipation for strong signal transfer: In a similar way to the design of fighter-jets, unstable systems can be easily changed by small inputs, which leads to more efficient signal transfer.



\textit{Ferromagnetic Resonance.}--- FMR is the phenomenon where the spin will follow a constant precession in a rotating external magnetic field. The angle $\theta$ from the $z$-axis at which it will do so in the steady state will vary according to the driving frequency $\omega_d$ of the magnetic field. Close to the natural frequency of the precession, one generally finds a resonance peak~\cite{Ludwig2020}. We assume a magnetic field of the form
\begin{equation}
	\bm{B}_{\text{eff}}(t)= \begin{pmatrix}
		\Omega \cos(\omega_d t)\\ \Omega \sin(\omega_d t)\\B_0 -K S_z
	\end{pmatrix},
\end{equation}
where $\Omega$ is the strength of the rotating component, and we will neglect thermal noise. We search for a steady state solution of $\bm{S}(t)$ in the rotating frame where $\bm{B}_{\text{eff}}(t)$ is constant. We will assume a small $\theta$ approximation where the ground state is in the positive $z$ direction, i.e.  $0<\Omega \ll B_0-K S$ and $\alpha_s S \ll (B_0-K S)^{1-s}$. Then (see Sup. Mat.~\cite{sm} for details of the calculations), we find that the resonance occurs at a driving frequency
 \begin{equation}
	\omega_\text{res} 
	\approx
	(B_0-K S)+ (B_0-K S)^s \alpha_s S  \cos\left(\frac{\pi s}{2}\right).\label{eq:kres}
\end{equation}
It should be noted that this is different from what was to be expected from any scaling arguments, since the cosine term is completely new compared to previous results~\cite{Ludwig2020}, and it vanishes precisely when $s=1$. However, this new non-linear term scales as $(B_0-K S)^s$, which is an easily controllable parameter. In the limit where $B_0-K S$ is small (resp. large), the linear term will vanish and the $s$-power scaling can be measured for the sub(resp. super)-Ohmic case.
The amplitude at resonance is found to be
\begin{equation}
		\sin^2\theta_\text{res} \approx \frac{\Omega^2}{\left[\alpha_s S (B_0-K S)^s \sin\left(\frac{\pi s}{2}\right)\right]^2}, \label{eq:amp}
\end{equation}
and the Full Width at Half Maximum (FWHM) linewidth is given by
\begin{figure}[t]
	\center
	\includegraphics[width=\columnwidth]{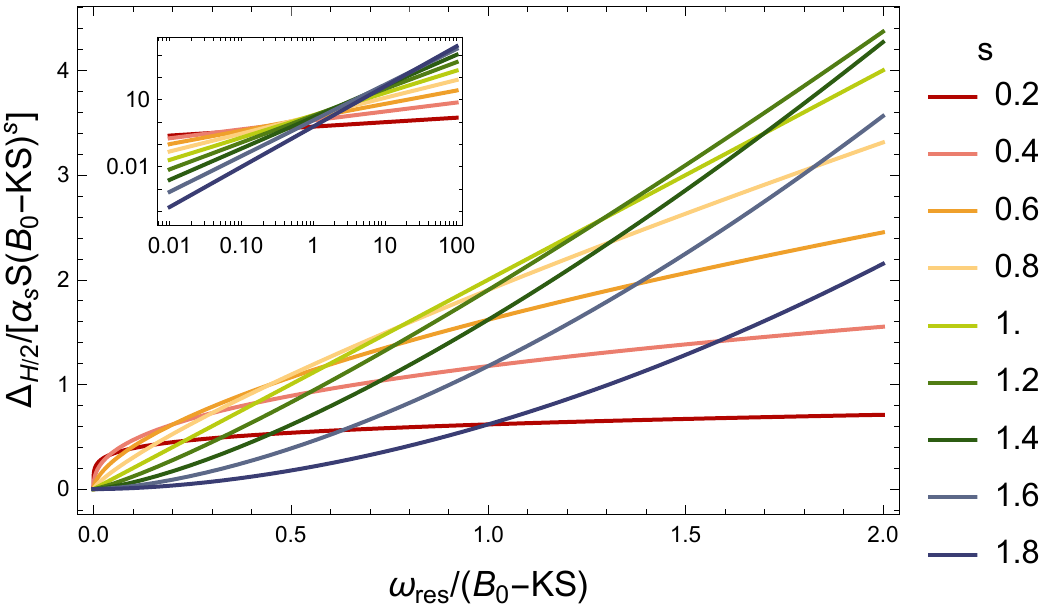}
	\caption{A plot of the linewidth in Eq.~\eqref{eq:fwhm2} as a function of resonance frequency for several values of $s$. The inset shows the same plot in a log-log scale, where the slope of the linewidth is precisely the Ohmicness $s$ of the bath.
	}
	\label{fig:fmr2}
\end{figure}
\begin{equation}
	 \Delta_{H/2} \approx 2\alpha_s S (B_0-K S)^s \sin\left(\frac{\pi s}{2}\right).\label{eq:fwhm}
\end{equation}
Depending on the experimental setup, it might be easier to measure either the resonance location or the width of the peak. Nevertheless, both will give the opportunity to see the $s$ scaling in $B_0-K S$. 
The presence of the anisotropy provides a good opportunity to reach weak or strong field limits. In fact, the orientation of the anisotropy can help to add or subtract from the magnetic field, which should make the required field strengths more reachable for experiments. Some setups are more suitable for measuring the width as a function of resonance frequency. When $s=1$, this relation can be directly derived from Eqs.~\eqref{eq:kres} and~\eqref{eq:fwhm}. However, when $s\neq 1$, the relation can only be approximated for strong or weak damping. For small $\alpha_s S$, we see that 
\begin{equation}
	\Delta_{H/2} \approx 2\alpha_s S (\omega_{\text{res}})^s \sin\left(\frac{\pi s}{2}\right).\label{eq:fwhm2}
\end{equation}

 The resonance peaks have been calculated numerically in FIG.~\ref{fig:fmr1} in dimensionless values. The red dashed lines show the location of the numerically calculated peak and the FWHM line width. The black solid lines show the location of the analytically approximated result for the peak location and FWHM line width [Eqs.~\eqref{eq:kres} and~\eqref{eq:fwhm}]. For small $\alpha_s S$ and $\Omega$, we see a good agreement between the analytical results and the numerical ones, although sub-Ohmic seems to match more closely than super-Ohmic. This could be due to the greater stability of sub-Ohmic systems, since the approximations might affect less a stable system. As one might expect from the thought experiment presented earlier, we can see in FIG.~\ref{fig:fmr1} that sub-Ohmic systems require higher, more energetic, driving frequencies to resonate, whereas super-Ohmic systems already resonate at lower, less energetic, driving frequencies. In FIG.~\ref{fig:fmr2}, we provide a  plot of Eq.~\eqref{eq:fwhm2} to facilitate further comparison with experiments. If the assumption of Gilbert damping was correct, all that one would see is a slope of one in the log-log inset.



\textit{Conclusion.}--- By relaxing the Ohmic Gilbert damping assumption, we have shown that the low-frequency regime of magnetization dynamics can be modeled by a fractional LLG equation. This was done by coupling the macrospin to a bath of harmonic oscillators in the framework of a Caldeira-Leggett model. The Keldysh formalism was used to compute the out-of-equilibrium dynamics of the spin system. By analyzing an FMR setup, we found an $s$-power scaling law in the resonance frequency and linewidth of the spin, which allows for a new way to  measure the value of $s$. This means that experiments in magnetization dynamics and spintronics can now avoid the assumption of Gilbert damping and instead measure the Ohmicness of the environment. This could aid in a better understanding of how to improve efficiency, stability, and control of such systems for practical applications.



\textit{Acknowledgments.}--- This work was supported by
the Netherlands Organization for Scientific Research (NWO,
Grant No. 680.92.18.05, C.M.S. and R.C.V.) and (partly) (NWO, Grant No. 182.069, T.L. and R.A.D.).


\bibliographystyle{apsrev}
\bibliography{manuscript}

\begin{thebibliography}{44}
\bibitem[{foo({\natexlab{b}})}]{footnote2}
\bibinfo{note}{The dissipation kernel $ \alpha$ is closely related to the
  retarded $\alpha^R$ and advanced $\alpha^A$ components. Namely, it is given
  by $ \alpha (\omega) = -\alpha^R(\omega) - \alpha^A(-\omega)$.}
\bibitem[{foo({\natexlab{c}})}]{footnote3}
\bibinfo{note}{From a mathematical perspective, any continuous but not smooth
  function can still be expanded in a power-law for small enough parameters.
  Hence, the only assumption that we make in this model is that the frequencies
  in the system are very small. Then, there will always exist an
  $0<s\in\mathbb{R}$ such that this expansion holds. In contrast, the Ohmic
  expansion can only be made for smooth continuous functions.}
\bibitem[{foo({\natexlab{d}})}]{footnote4}
\bibinfo{note}{This definition does not have any boundary conditions, as they
  would have to be at $-\infty$ and would dissipate before reaching a finite
  time. One can, however, enforce boundary conditions by applying a very strong
  magnetic field for some time such that the spin aligns itself, and then
  quickly change to the desired field at $t=0$.}
\end{thebibliography}

\clearpage

\end{document}


\title		{The fractional Landau-Lifshitz-Gilbert equation\\Supplementary Material}
\author		{R.C. Verstraten$^{1}$, T. Ludwig$^{1}$, R.A. Duine$^{1,2}$, C. Morais Smith$^{1}$}
\affiliation	{$^{1}$Institute for Theoretical Physics, Utrecht University, Princetonplein 5, 3584CC Utrecht, The Netherlands \\
$^{2}$Department of Applied Physics, Eindhoven University of Technology,
P.O. Box 513, 5600 MB Eindhoven, The Netherlands}
\date{\today}

%



\maketitle

\tableofcontents

\section{Keldysh microscopic model}
For pedagogical reasons we start with a microscopic derivation of the usual LLG equation before going into the fractional one.
In this section, we combine spin coherent states with the Keldysh formalism~\cite{altlandsimons2010,Kamenev2011} to derive a stochastic Langevin-like equation of motion of a (macro) spin~\cite{schmid1982quasiclassical}.

\subsection{Hamiltonian}
In the main text, we introduced a spectral function $J(\omega)$ with a cutoff frequency $\Omega_c$. This was originally done from the perspective that any spectral function could be expanded to linear order; hence, the model would only be valid up to some highest frequency. However, the cutoff is also important for the model to be realistic, since any physical spectral function should vanish as $\omega\to \infty$. In the main text, we stated that the same results can be obtained by introducing a constant counter term in the Hamiltonian. This is a term which exactly completes the square of the coupling term and the harmonic potential of the bath and can be seen as a normalization of the zero-energy level. If we instead start the model with this counter term and drop the cutoff, we will get a Greens function $\alpha_\text{ct}(\omega)$, which is precisely such that the original Greens function can be written as $\alpha(\omega)=\alpha(0)+ \alpha_\text{ct}(\omega)$, i.e., the counter term in the Hamiltonian removes the zero frequency contribution of the Greens function. This $\alpha(\omega=0)$ generates a term in the equation of motion that goes as $\int_0^\infty d\epsilon \frac{J(\epsilon)}{\pi \epsilon} [\bm{S}(t)\times\bm{S}(t)]$. Since the integral is finite, with a frequency cutoff in $J(\omega)$, the entire term is zero due to the cross product. This means that the equation of motion will be identical if we start either from the regular Hamiltonian with a frequency cutoff, or with a counter term and no cutoff. Here, we choose to show the method that includes a counter term, because then we do not need to calculate terms which would have canceled either way.

The microscopic system that we describe is a large spin in an external magnetic field, where the spin is linearly coupled to a bath of harmonic oscillators in the same way as in Refs.~\cite{Caldeira1981,Caldeira1983physa,Caldeira1983annphys,Caldeira2012,caldeira1985influence,Weiss2012}. Therefore, our Hamiltonian has the form of a system, coupling, bath, and counter term; $H(t) = H_s +H_c +H_b +H_{ct}$, where
\begin{align}
	H_s &= \bm{B}\cdot\hat{\bm{S}}-K S_z^2,\nonumber\\
	H_c &=  \sum_\alpha \gamma_\alpha \hat{\bm{S}} \cdot\hat{\bm{x}}_\alpha,\nonumber\\
	H_b &= \sum_\alpha \frac{\hat{\bm{p}}_\alpha^2}{2m_\alpha} +\frac{m_\alpha \omega_\alpha^2}{2} \hat{\bm{x}}_\alpha^2,\nonumber\\
	H_{ct} &= \sum_\alpha \frac{ \gamma_\alpha^2}{2m_\alpha \omega_\alpha^2} \hat{\bm{S}}^2.
\end{align}
Here, $\bm{B}$ is the (effective) magnetic field, $\hat{\bm{S}}$ is the spin, $K$ is the $z$-axis anisotropy, $\gamma_\alpha$ is the coupling strength, and $\alpha$ is the index over all harmonic oscillators which have position $\hat{\bm{x}}_\alpha$, momentum $\hat{\bm{p}}_\alpha$, mass $m_\alpha$ and natural frequency $\omega_\alpha$. Notice that the counter term is constant, since $S^2$ is a conserved quantity, and that we have indeed completed the square, such that 
\begin{align}
	H(t)=\bm{B}\cdot\hat{\bm{S}}-K S_z^2 +\sum_\alpha \frac{\hat{\bm{p}}_\alpha^2}{2m_\alpha} + \left[ \sqrt{\frac{m_\alpha \omega_\alpha^2}{2}} \hat{\bm{x}}_\alpha+  \sqrt{\frac{ \gamma_\alpha^2}{2m_\alpha \omega_\alpha^2}}\hat{\bm{S}} \right]^2.
\end{align}

\subsection{Keldysh partition function}
We will use the Keldysh formalism to derive a quasi-classical equation of motion. Since this is an out-of-equilibrium system, a common choice would be to use the Lindblad formalism with a master equation~\cite{gardiner2004quantum}. However, Lindblad can only describe Markovian systems, which will not be the case when we introduce a non-Ohmic bath. In the Keldysh formalism, one starts with an equilibrium density matrix in the far past (effectively infinite on the relevant time scale). This then gets evolved with the time evolution operator as usual. However, in contrast to ordinary path integrals, once the present has been reached, one evolves back to the infinite past. Since there is infinite time for evolution, we can reach out-of-equilibrium states adiabatically. The benefit of integrating back to the infinite past is that we begin and end with the same in-equilibrium system, which means equilibrium techniques can be used, at the cost of having both the forward ($\mathcal{O}^+$) and backward ($\mathcal{O}^-$) quantities to take care of. To reach useful results, one can apply a Keldysh rotation to the classical ($	\mathcal{O}^c= (\mathcal{O}^+ +\mathcal{O}^-)/2 $) and quantum ($ \mathcal{O}^q = \mathcal{O}^+ -\mathcal{O}^-$) components with the added notation $\vec{\mathcal{O}}= \begin{pmatrix}
	\mathcal{O}^c\\\mathcal{O}^q \end{pmatrix}$. To derive a quasi-classical equation of motion, the action can be expanded in all the quantum components, after which the Euler-Lagrange equation for the quantum components provides the equation of motion in terms of the classical components.

\begin{figure}[h]
	\center
	\includegraphics[width=0.8\linewidth]{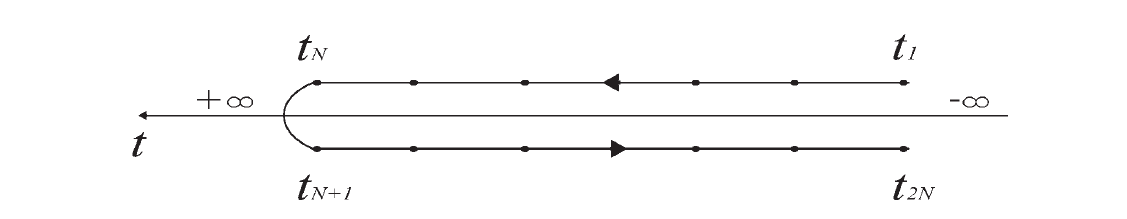}
	\caption{Figure extracted from Ref.~\cite{Kamenev2011}. The Keldysh contour starts at $t=-\infty$, evolves forward to some time $t$, and then evolves backwards in time to $t=-\infty$.}
	\label{fig:keldyshcontour}
\end{figure}

To begin, we write down the Keldysh partition function
\begin{align}
	Z&= \Tr\left\{ T_K \exp\left[-i \oint_K dt H(t)\right] \rho_0 \right\},
\end{align}
where $T_K$ is the Keldysh time ordering, $\rho_0$ is the density matrix at $t=-\infty$, and the integral runs over the Keldysh contour, as shown in FIG.~\ref{fig:keldyshcontour}. After discretizing the Keldysh time integral in the way of FIG.~\ref{fig:keldyshcontour}, we can rewrite the trace as path-integrals over the spin coherent state $\ket{g}$ and the oscillators $\ket{\hat{\bm{x}}_\alpha}$ and $ \ket{\hat{\bm{p}}_\alpha}$. This yields
\begin{equation}
	Z= \int Dg \prod_\alpha \int D\hat{\bm{x}}_\alpha \int D\hat{\bm{p}}_\alpha \;e^{i S[g,\{\hat{\bm{x}}_\alpha\},\{\hat{\bm{p}}_\alpha\}]},
\end{equation}
with the Keldysh action
\begin{align}
	 S[g,\{\hat{\bm{x}}_\alpha\},\{\hat{\bm{p}}_\alpha\}]= 
	\oint_K dt& \Bigg[ (-i\partial_t \bra{g})\ket{g} -\bm{B}\cdot \bm{S}_g +K S_{z,g}^2 \nonumber\\
	&+\left.\sum_\alpha \left(
	-\gamma_\alpha  \bm{S}_g \cdot \hat{\bm{x}}_\alpha 
	+\hat{\bm{p}}_{\alpha}\cdot\dot{\hat{\bm{x}}}_\alpha -\frac{ \gamma_\alpha^2\bm{S}_g^2}{2m_\alpha \omega_\alpha^2} 
	-\frac{\hat{\bm{p}}_\alpha^2}{2m_\alpha} -\frac{m_\alpha \omega_\alpha^2}{2}\hat{\bm{x}}_\alpha^2
	\right) 	   \right],
\end{align}
where we defined $\bm{S}_g = \bra{g}\hat{\bm{S}}\ket{g}$.

The continuous path-integral seems to miss the boundary term $\bra{\hat{\bm{x}}_{1,\alpha},g_{1}} \rho_0 \ket{\hat{\bm{x}}_{2N,\alpha},g_{2N}}\braket{\hat{\bm{p}}_{2N,\alpha}}{\hat{\bm{x}}_{2N,\alpha}}$, but it is included in the Keldysh contour, as it connects the beginning and final contour time at $t=-\infty$; see Ref.~\cite{Kamenev2011}.

Now, we will integrate out the bath degrees of freedom, beginning by completing the square and performing the Gaussian integral over $\hat{\bm{p}}_\alpha$. The Gaussian contribution in $\hat{\bm{p}}_\alpha$ will act as a constant prefactor, so it will drop out of any calculation of an observable due to the normalization. Hence, we can effectively set it to one to find
\begin{align}
	\int D\hat{\bm{p}}_\alpha \exp\left[-i\oint_K dt \left( \frac{\hat{\bm{p}}_\alpha^2}{2m_\alpha} -\hat{\bm{p}}_{\alpha}\cdot\dot{\hat{\bm{x}}}_\alpha \right) \right]&
	=
	\exp\left[i\oint_K dt \left(    -\frac{m_\alpha}{2}\hat{\bm{x}}_\alpha\partial_t^2\hat{\bm{x}}_\alpha \right) \right],
\end{align}
where we also did a partial integration in $\hat{\bm{x}}_\alpha$. Next we will perform a similar approach for the positions, but it is useful to apply the Keldysh rotation first. Note that we can directly rewrite the integral over the Keldysh contour as a regular time integral over the quantum components. However, one must still rewrite the contents of the integral in terms of the quantum and classical parts of the variables, since the Keldysh rotation does not immediately work for products. The action can first be written as
\begin{align}
	i S[g,\{\hat{\bm{x}}_\alpha\}]&=  
	i\int dt \left( \left[ (-i\partial_t \bra{g})\ket{g} \right]^q -\left[ \bm{B}\cdot \bm{S}_g \right]^q +K\left[  S_{z,g}^2\right]^q \right.\nonumber\\
	&-\sum_\alpha\left. \left\{\left[ 
	\gamma_\alpha  \bm{S}_g \cdot \hat{\bm{x}}_\alpha \right]^q +\frac{ \gamma_\alpha^2[\bm{S}_g^2]^q}{2m_\alpha \omega_\alpha^2}+ \left[ \frac{m_\alpha}{2}\hat{\bm{x}}_\alpha \left( \partial_t^2
	+ \omega_\alpha^2\right)\hat{\bm{x}}_\alpha \right]^q
	\right\}  \right).
\end{align}
We can then derive that
\begin{align}
-	\left[	\gamma_\alpha  \bm{S}_g \cdot \hat{\bm{x}}_\alpha\right]^q &= -\gamma_\alpha\left[ \bm{S}_g^+ \cdot \hat{\bm{x}}_\alpha^+ -\bm{S}_g^- \cdot \hat{\bm{x}}_\alpha^- \right] = -\gamma_\alpha\left[ \left(\bm{S}_g^c +\frac{1}{2}\bm{S}_g^q\right) \cdot \left(\hat{\bm{x}}_\alpha^c +\frac{1}{2}\hat{\bm{x}}_\alpha^q\right) -\left(\bm{S}_g^c -\frac{1}{2}\bm{S}_g^q\right) \cdot \left(\hat{\bm{x}}_\alpha^c -\frac{1}{2}\hat{\bm{x}}_\alpha^q\right) \right] \nonumber\\ &=-\gamma_\alpha\left[ \bm{S}_g^c\hat{\bm{x}}_\alpha^q +\bm{S}_g^q\hat{\bm{x}}_\alpha^c \right]= -\gamma_\alpha\left[ \begin{pmatrix}\bm{S}_g^c & \bm{S}_g^q\end{pmatrix} 
	\tau_{x}
	\begin{pmatrix} \hat{\bm{x}}_\alpha^c\\\hat{\bm{x}}_\alpha^q 	\end{pmatrix}
	\right], \label{eq:qcompsx}
\end{align}
where we introduced $\tau_{x}= \begin{pmatrix}  0 &1\\1&0	\end{pmatrix} $ in the Keldysh (classical, quantum) space represented by an upper index $c$ and $q$ respectively. Next, we want to derive a similar form for the part of the action that is quadratic in $\hat{\bm{x}}_\alpha$. Since these are harmonic oscillators in equilibrium, we can refer the reader to Ref.~\cite{Kamenev2011}, noting that a unit mass was used there, and conclude that 
\begin{align}
	\left[	- \frac{m_\alpha}{2}\hat{\bm{x}}_\alpha \left( \partial_t^2
	+ \omega_\alpha^2\right)\hat{\bm{x}}_\alpha \right]^q &=
	\begin{pmatrix} \hat{\bm{x}}_\alpha^c &\hat{\bm{x}}_\alpha^q\end{pmatrix} 
	\begin{pmatrix} 0 & \left[ G_\alpha^{-1}\right]^A \\ \left[ G_\alpha^{-1}\right]^R &\left[ G_\alpha^{-1}\right]^K	\end{pmatrix}
	\begin{pmatrix} \hat{\bm{x}}_\alpha^c\\\hat{\bm{x}}_\alpha^q 	\end{pmatrix},
\end{align}
where the retarded and advanced Greens functions read
\begin{equation}
	[G^{-1}_\alpha]^{R/A}(t-t') = \delta(t-t') \frac{m_\alpha}{2}[(i\partial_t \pm i0)^2 -\omega_\alpha^2].
\end{equation} 
The $\pm i0$ is introduced because we need an infinitesimal amount of dissipation on the bath for it to remain in equilibrium and the sign is tied to causality. This is because there is also an infinitesimal amount of energy transfer from the macroscopic spin to each of the oscillators. This results in an extra first-order derivative term, which is found by multiplying out the square with $i0$. One might want to set these terms to zero immediately, but as it turns out, these are very important limits, which shift away poles from integrals that we need to compute later. Once that is done, the limits are no longer important for the final result, and they may finally be put to zero. Since the bath is in equilibrium, we can use the fluctuation dissipation theorem to compute the Keldysh component using
\begin{equation*}
	G^K_\alpha(\omega)=\left[G^R_\alpha(\omega) -G^A_\alpha(\omega)\right]\coth\left(\frac{\omega}{2T}\right).
\end{equation*}
The $\hat{\bm{x}}$ dependent part of the action is now given by 
\begin{align}
	i S_X
	&= i \int dt \left[ -\gamma_\alpha \begin{pmatrix}\bm{S}_g^c & \bm{S}_g^q\end{pmatrix} 
	\tau_{x}
	\begin{pmatrix} \hat{\bm{x}}_\alpha^c\\\hat{\bm{x}}_\alpha^q 	\end{pmatrix}
	+
	\begin{pmatrix} \hat{\bm{x}}_\alpha^c &\hat{\bm{x}}_\alpha^q\end{pmatrix} 
	G_\alpha^{-1}
	\begin{pmatrix} \hat{\bm{x}}_\alpha^c\\\hat{\bm{x}}_\alpha^q 	\end{pmatrix} \right],\label{eq:SX}
\end{align}
which we can compute by completing the square to find
\begin{equation}
	i S_X= i \int dt \left[-\frac{\gamma_\alpha^2}{4} \vec{S}^T_g \begin{pmatrix} 0 & G_\alpha^A\\ G_\alpha^R & G_\alpha^K \end{pmatrix} \vec{S}_g\right].
\end{equation}
Before we write down the final effective action, we also have to rewrite the quadratic part in $\bm{S}$ in a similar vector form, which is
\begin{align}
	-\frac{ \gamma_\alpha^2[\bm{S}_g^2]^q}{2m_\alpha \omega_\alpha^2} &
	= \frac{ -\gamma_\alpha^2}{2m_\alpha \omega_\alpha^2} \begin{pmatrix}\bm{S}_g^c & \bm{S}_g^q\end{pmatrix} 
	\tau_{x}
	\begin{pmatrix} \bm{S}_g^c \\ \bm{S}_g^q	\end{pmatrix}.
\end{align}	
Combining everything together, we find that the partition function of the system is given by $	Z= \int Dg  \;e^{i S[g]}$, 
with the effective action
\begin{align}
	i S[g]&= i \int dt \left\{ \left[ (-i\partial_t \bra{g})\ket{g} -\bm{B}\cdot \bm{S}_g +K S_{z,g}^2 \right]^q - \int dt' \vec{S}_g^T(t) \begin{pmatrix} 0 & \alpha^A\\ \alpha^R & \alpha^K \end{pmatrix}_{(t-t')} \vec{S}_g(t')\right\}, \label{eq:krots}
\end{align}
where $\alpha^{A/R}(t-t') =   \sum_\alpha \left( \frac{\gamma_\alpha^2}{4} G_\alpha^{A/R}(t-t') +\frac{ \gamma_\alpha^2}{2m_\alpha \omega_\alpha^2}\delta(t-t')\right)$ and  $\alpha^{K}(t-t') =   \sum_\alpha \frac{\gamma_\alpha^2}{4} G_\alpha^{K}(t-t') $.

\subsection{Quasi-classical equation of motion}
In the quasi-classical regime, we are interested in solutions where the quantum components $(q)$ are small compared to the classical components $(c)$. We can thus neglect terms of $\mathcal{O}[(q)^3]$, but we must be careful with $(q)^2$. We can use a Hubbard-Stratonovich transformation to convert $(q)^2$ terms into an expression with just $(q)$, but with a new field $\xi$ added to the path integral~\cite{schmid1982quasiclassical}. The action will then contain only terms of linear order in $(q)$, which means the partition function has the form $Z\sim \int Dc Dq \exp[i f(c) q]= \int Dc \frac{1}{2\pi} \delta[f(c)]$.  Hence, only solutions that satisfy $f(c)=0$ contribute to the path integral. Within that subset, we want to minimize the action.

In order to derive the equation of motion of the system, we must understand the relation between $\ket{g}$ and $\bm{S}_g =\bra{g}\bm{S}\ket{g}$. Using the Euler angle representation \cite{altlandsimons2010}, we can describe $\ket{g}$ as 
\begin{equation}
	\ket{g}= g\ket{\uparrow} = e^{-i\phi S_z} e^{-i\theta S_y} e^{-i\psi S_z} \ket{\uparrow}= e^{-i\phi S_z} e^{-i\theta S_y}  \ket{\uparrow} e^{-i\psi S}
\end{equation}
and similarly
\begin{equation}
	\bra{g}=  e^{i\psi S} \bra{\uparrow} e^{i\theta S_y}  e^{i\phi S_z}.
\end{equation}
Note that the $\psi$ angle is now independent of the quantum state $\ket{\uparrow}$, since this angle is describing the rotation of the vector pointing in the spin direction, which is symmetric. Hence, this will yield a gauge symmetry.

Using the Euler angle representation in the first terms of Eq. \eqref{eq:krots}, we see that
\begin{align}
	(-i\partial_t \bra{g} )\ket{g} &=
	\left( \dot{\psi}Se^{i\psi S} \bra{\uparrow} e^{i\theta S_y}  e^{i\phi S_z} + e^{i\psi S} \bra{\uparrow} \dot{\theta} S_y e^{i\theta S_y} e^{i\phi S_z} + e^{i\psi S} \bra{\uparrow} e^{i\theta S_y} \dot{\phi} S_z e^{i\phi S_z}  \right)e^{-i\phi S_z} e^{-i\theta S_y}  \ket{\uparrow} e^{-i\psi S}\nonumber\\
	&=
	\dot{\psi}S +  \dot{\theta} \bra{\uparrow} S_y \ket{\uparrow} + \dot{\phi} \bra{\uparrow} e^{i\theta S_y}  S_z e^{-i\theta S_y}\ket{\uparrow}.
\end{align}
We note that $\bra{\uparrow} S_y \ket{\uparrow}=0$, while the last term includes a rotation of the spin up state by $\theta$ degrees in the $y$ direction and then measures the $S_z$ component of that state, which is $S \cos \theta$. Hence, 
\begin{equation}
	(-i\partial_t \bra{g} )\ket{g} = \dot{\psi}S + \dot{\phi} S \cos \theta.
\end{equation}
We now define a new variable $\chi$ such that $\psi=\chi-\phi$, which results in 
\begin{equation}
	(-i\partial_t \bra{g} )\ket{g} = \dot{\chi}S - \dot{\phi} (1- \cos \theta)S. \label{eq:idtgg}
\end{equation}
Making use of the Euler angle representation, we also see that 
\begin{equation}
	\bm{S}_g= S \begin{pmatrix}
		\sin\theta \cos\phi \\ \sin\theta\sin\phi \\ \cos\theta
	\end{pmatrix}.
\end{equation}
We see that $\bm{B}\cdot\bm{S}_g= S[B_x \sin\theta\cos\phi+ B_y \sin\theta\sin\phi +B_z \cos\theta]$. Similarly, $K \bm{S}_{z,g}^2= K S^2 \cos^2\theta$.  Now, we still have to compute the quantum parts of these quantities. We first note that 
\begin{align}
	S_{g, x}^q/S &=[\sin\theta \cos\phi]_q =2\cos\theta_c\sin\frac{\theta_q}{2}\cos\phi_c\cos\frac{\phi_q}{2} -2\sin\theta_c\cos\frac{\theta_q}{2}\sin\phi_c\sin\frac{\phi_q}{2};\nonumber\\
	S_{g, y}^q/S&=[\sin\theta\sin\phi ]_q =2\sin\theta_c\cos\frac{\theta_q}{2}\cos\phi_c\sin\frac{\phi_q}{2} +2\cos\theta_c\sin\frac{\theta_q}{2}\sin\phi_c\cos\frac{\phi_q}{2};\nonumber\\
		S_{g, z}^q/S&=[\cos\theta]_q= -2\sin\theta_c\sin\frac{\theta_q}{2};\nonumber\\
		[\cos^2\theta]_q&=-2\sin\theta_c\cos\theta_c \sin\theta_q.\label{eq:sangle}
\end{align}
 Next, we will choose a gauge for $\chi$ as in Ref. \cite{Shnirman2015}, which is 
\begin{align}
	\dot{\chi}_c &= \dot{\phi}_c(1-\cos \theta_c)\nonumber\\
	\chi_q &= \phi_q (1-\cos\theta_c).
\end{align}
Defining $p=1-\cos\theta$, we see that $	\left[ 	(-i\partial_t \bra{g} )\ket{g} \right]_q =\left[ \dot{\chi}S - \dot{\phi} p S \right]_q =S\left[  \phi_q \dot{p}_c - \dot{\phi}_c p_q \right]$. Now, 
$p_q= 2\sin\theta_c\sin\frac{\theta_q}{2}$ and 
$\dot{p}_c = \dot{\theta}_c \sin\theta_c \cos\frac{\theta_q}{2} + \frac{\dot{\theta}_q}{2} \cos\theta_c \sin\frac{\theta_q}{2},$
which leads to 
\begin{align}
	\left[ 	(-i\partial_t \bra{g} )\ket{g} \right]_q 
	&=S\left[  \phi_q \dot{p}_c - \dot{\phi}_c p_q \right]
	= S\left[  \phi_q \dot{\theta}_c \sin\theta_c \cos\frac{\theta_q}{2} + \phi_q\frac{\dot{\theta}_q}{2} \cos\theta_c \sin\frac{\theta_q}{2} -2 \dot{\phi}_c \sin\theta_c\sin\frac{\theta_q}{2} \right].
\end{align}
Next, we want to express $\bm{B}\cdot\bm{S}_g^q$ in terms of Euler angles. We see that
\begin{align}
	\bm{B}\cdot\bm{S}_g^q &= S[B_x \sin\theta\cos\phi+ B_y \sin\theta\sin\phi +B_z \cos\theta]^q \nonumber\\
	&= 2S\Big[ B_x\left(\cos\theta_c\sin\frac{\theta_q}{2}\cos\phi_c\cos\frac{\phi_q}{2} -\sin\theta_c\cos\frac{\theta_q}{2}\sin\phi_c\sin\frac{\phi_q}{2}\right)\nonumber\\ &\qquad+B_y\left(\sin\theta_c\cos\frac{\theta_q}{2}\cos\phi_c\sin\frac{\phi_q}{2} +\cos\theta_c\sin\frac{\theta_q}{2}\sin\phi_c\cos\frac{\phi_q}{2}\right) \nonumber\\
	&\qquad-B_z\sin\theta_c\sin\frac{\theta_q}{2} \Big],
\end{align}
where we used the results from Eq.~\eqref{eq:sangle}. Similarly, we have
\begin{align}
	K \left[S_{z,g}^2\right]^q &=K S^2 [\cos^2\theta]^q= -2K S^2\sin\theta_c\cos\theta_c \sin\theta_q.
\end{align}
Combining these results, we conclude that
\begin{align}
	\left[ (-i\partial_t \bra{g})\ket{g} -\bm{B}\cdot \bm{S}_g +K S_{z,g}^2 \right]^q =S\Bigg[ & \phi_q \dot{\theta}_c \sin\theta_c \cos\frac{\theta_q}{2} + \phi_q\frac{\dot{\theta}_q}{2} \cos\theta_c \sin\frac{\theta_q}{2} -2 (-B_z+K S\cos\theta_c +\dot{\phi}_c )\sin\theta_c\sin\frac{\theta_q}{2}\nonumber\\
	 &-2B_x\left(\cos\theta_c\sin\frac{\theta_q}{2}\cos\phi_c\cos\frac{\phi_q}{2} -\sin\theta_c\cos\frac{\theta_q}{2}\sin\phi_c\sin\frac{\phi_q}{2}\right)\nonumber\\ &-B_y\left(\sin\theta_c\cos\frac{\theta_q}{2}\cos\phi_c\sin\frac{\phi_q}{2} +\cos\theta_c\sin\frac{\theta_q}{2}\sin\phi_c\cos\frac{\phi_q}{2}\right)\Bigg]. \label{eq:gpart}
\end{align}
Remark that this expression only contains odd powers of $(q)$, so that we can neglect all higher-order terms to get
\begin{align}
	\Big[ (-i\partial_t \bra{g})\ket{g} &-\bm{B}\cdot \bm{S}_g +K \bm{S}_{z,g}^2\Big]^q =S\Big[   -\theta_q\sin\theta_c (-B_z+KS\cos\theta_c +\dot{\phi}_c ) \nonumber\\  
	&-\theta_q\cos\theta_c(B_x\cos\phi_c +B_y\sin\phi_c) +\phi_q\sin\theta_c(\dot{\theta}_c +B_x\sin\phi_c- B_y\cos\phi_c )	 \Big]. \label{eq:gpartlinq}
\end{align}

Now, we focus on the part of the action in Eq. \eqref{eq:krots} that comes from the bath, given by
\begin{equation}
	i S_{b}[g]= -i \int dt\int dt' \vec{S}_g^T(t) \begin{pmatrix} 0 & \alpha^A\\ \alpha^R & \alpha^K \end{pmatrix}_{(t-t')} \vec{S}_g(t').
\end{equation}
Let us first consider what $\bm{S}_g^q$ and $\bm{S}_g^c$ are in terms of $\phi$ and $\theta$. By performing some trigonometric operations on each of the components, we find that	
\begin{align}
	\bm{S}_g^c	&= S \begin{pmatrix}
		\sin\theta_c\cos\frac{\theta_q}{2}\cos\phi_c\cos\frac{\phi_q}{2} -\cos\theta_c\sin\frac{\theta_q}{2}\sin\phi_c\sin\frac{\phi_q}{2} \\ 
		\sin\theta_c\cos\frac{\theta_q}{2}\sin\phi_c\cos\frac{\phi_q}{2} +\cos\theta_c\sin\frac{\theta_q}{2}\cos\phi_c\sin\frac{\phi_q}{2}\\
		\cos\theta_c\cos\frac{\theta_q}{2}
	\end{pmatrix} 
\end{align}
and
\begin{align}
	\bm{S}_g^q	&= 2S \begin{pmatrix}
		\cos\theta_c\sin\frac{\theta_q}{2}\cos\phi_c\cos\frac{\phi_q}{2} -\sin\theta_c\cos\frac{\theta_q}{2}\sin\phi_c\sin\frac{\phi_q}{2} \\ 
		\sin\theta_c\cos\frac{\theta_q}{2}\cos\phi_c\sin\frac{\phi_q}{2} +\cos\theta_c\sin\frac{\theta_q}{2}\sin\phi_c\cos\frac{\phi_q}{2}\\
		-\sin\theta_c\sin\frac{\theta_q}{2}
	\end{pmatrix} .
\end{align}
By expanding in the quantum components of $\bm{S}_g^c$ and $\bm{S}_g^q$, we see that 
\begin{align*}
	\bm{S}_g^c = (q)^0 +\mathcal{O}\left((q)^2\right), \\ 
	\bm{S}_g^q = (q)^1 +\mathcal{O}\left((q)^3\right).
\end{align*}
Since the action only contains terms with at least one $\bm{S}_g^q$, we know that the only way to obtain a term of order $(q)^2$ is from $(\bm{S}_g^q)^2$. Hence, we may neglect all terms beyond linear $(q)$ in $\bm{S}_g^{\;(c/q)}$ in the quasi-classical regime. This results in
\begin{align}
	\bm{S}_g^c	&= S \begin{pmatrix}
		\sin\theta_c\cos\phi_c  \\ 
		\sin\theta_c\sin\phi_c \\
		\cos\theta_c
	\end{pmatrix}, \\
	\bm{S}_g^q	&= S \begin{pmatrix}
		\theta_q\cos\theta_c\cos\phi_c -\phi_q\sin\theta_c\sin\phi_c \\ 
		\phi_q\sin\theta_c\cos\phi_c +\theta_q\cos\theta_c\sin\phi_c\\
		-\theta_q\sin\theta_c
	\end{pmatrix} .
\end{align}
A useful remark for later is that this shows that 
\begin{equation}
	\bm{S}_g^q= \theta_q\frac{\partial}{\partial \theta_c} \bm{S}_g^c + \phi_q\frac{\partial}{\partial \phi_c} \bm{S}_g^c.\label{eq:qcrelation}
\end{equation}
Going back to $i S_{b}[g]$, we can rewrite this as a convolution, in the sense that
\begin{equation}
	iS_{b}[g] =-i \int dt\;\left[ \bm{S}_g^c(t) \cdot\left(\alpha^A * \bm{S}_g^q \right)(t)
	+\bm{S}_g^q(t) \cdot\left(\alpha^R * \bm{S}_g^c \right)(t)+
	\bm{S}_g^q(t) \cdot\left(\alpha^K * \bm{S}_g^q \right)(t)\right],
\end{equation}
where $(f*g)(t)=\int_{-\infty}^\infty dt' f(t-t')g(t')$. We see that the first two terms contain precisely one quantum component, but the last term has two quantum components. When writing down the Euler-Lagrange equation of motion, it is important to realize that the convolution operation will act as if it is a simple multiplication, since the convolution obeys
\begin{equation}
	\frac{d}{dx}(f(x)*g)(t)= \left(\frac{df}{dx}*g\right)(t).
\end{equation} 
We now concentrate on the $(q)^2$ part of this action, for which we would like to use a Hubbard-Stratonovich transformation in order to reduce this to linear in $(q)$. Recall that a Hubbard-Stratonovich transformation is given by
\begin{equation}
	\exp\left[-\frac{a}{2}x^2 \right] = \sqrt{\frac{1}{2\pi a}} \int D\xi \exp\left[ -\frac{\xi^2}{2a} -i x \xi \right].
\end{equation}
However, we see that our action does not contain any purely quadratic terms, but rather a Greens functional shape as $\bm{S}_g^q(t)\alpha^K(t-t')\bm{S}_g^q(t')$. Hence, to use a Hubbard-Stratonovich like transformation, we must derive it from a Greens function exponential, similarly to Ref.~\cite{schmid1982quasiclassical}. Assuming that this is renormalizable and that $\alpha^K$ can be rewritten into a distribution, we have 
\begin{align*}
	1&= \int D\xi \exp\left[ -\frac{1}{2} \int dt \int dt' \bm{\xi}(t) [-2i\alpha^K]^{-1}(t-t')\bm{\xi}(t') \right]\\
	&= \int D\xi \exp\left[ -\frac{1}{2} \int dt \int dt' \left(\bm{\xi}(t) -2 \int dt'' \bm{S}_g^q(t'')\alpha^K(t''-t)\right)[-2i\alpha^K]^{-1}(t-t')\left(\bm{\xi}(t') -2\int dt'''\alpha^K(t'-t''')\bm{S}_g^q(t''')\right) \right]\\ 
	&= \int D\xi \exp\left[ -\frac{1}{2} \int dt \int dt' \bm{\xi}(t) [-2i\alpha^K]^{-1}(t-t')\bm{\xi}(t')\right.\\
	&\qquad-\left.
	i\bm{S}_g^q(t) \delta(t-t')\bm{\xi}(t')
	-i\bm{\xi}(t) \delta(t-t')\bm{S}_g^q(t')
	-2i\bm{S}_g^q(t) \alpha^K(t-t')\bm{S}_g^q(t') \right]\nonumber\\
	&= \int D\xi \exp\left[ -\frac{1}{2} \int dt \int dt' \bm{\xi}(t) [-2i\alpha^K]^{-1}(t-t')\bm{\xi}(t')
	-2i\bm{S}_g^q(t) \delta(t-t')\bm{\xi}(t')
	-2i\bm{S}_g^q(t) \alpha^K(t-t')\bm{S}_g^q(t') \right],
\end{align*}
where we used that $\int dt' \alpha^K(t-t')[\alpha^K]^{-1}(t'-t'')= \delta(t-t'')$ and that $2i\alpha^K$ is positive real. Therefore, we find that
\begin{align}
	\exp\left[-i \int dt \int dt'  \bm{S}_g^q(t) \alpha^K(t-t')\bm{S}_g^q(t') \right]&= \int D\xi \exp\left[ -\frac{1}{2} \int dt \int dt'\bm{\xi}(t) [-2i\alpha^K]^{-1}(t-t')\bm{\xi}(t') \right] \nonumber\\
	&\qquad \cdot \exp\left[ i \int dt \bm{S}_g^q(t) \bm{\xi}(t) \right].
\end{align}
The double integral in the first exponential signifies the statistical properties of $\bm{\xi}$. For instance, if $ \alpha^K$ is delta-like, then $\bm{\xi}$ would have Gaussian statistics (e.g. white noise), but in general we will have time correlated noise defined by $\alpha^K$ \cite{schmid1982quasiclassical}, such that 
\begin{equation}
	\langle \bm{\xi}(t) \bm{\xi}(t') \rangle = -2i \alpha^K(t-t').
\end{equation}
Since there is no $g$ dependence in the double $\bm{\xi}$ exponential, we will leave it out of $S[g]$ and only remember these statistics. Our partition function is then given by
\begin{equation}
	Z= \int D\bm{\xi} \exp\left(i S_{n}\left[\bm{\xi}\right] \right) \int Dg \exp\left(iS_{sc}\left[g,\bm{\xi}\right]\right),
\end{equation}
where the noise action is given by 
\begin{equation}
	i S_{n}\left[\bm{\xi}\right] = -\frac{1}{2} \int dt \int dt'\bm{\xi}(t) [-2i\alpha^K]^{-1}(t-t')\bm{\xi}(t')
\end{equation}
and the semi-classical action is given by
\begin{align}
	iS_{sc}\left[g,\bm{\xi}\right] = i \int dt& \; S\Big[   -\theta_q\sin\theta_c (-B_z+KS\cos\theta_c +\dot{\phi}_c ) -\theta_q\cos\theta_c(B_x\cos\phi_c +B_y\sin\phi_c) \nonumber\\
	&+\phi_q\sin\theta_c(\dot{\theta}_c +B_x\sin\phi_c- B_y\cos\phi_c )	 \Big] +i \int dt\;\left[ \bm{\xi}(t)	\bm{S}_g^q(t)\right] \nonumber\\
	& - i \int dt\;\left[ \bm{S}_g^c(t) \cdot\left(\alpha^A * \bm{S}_g^q \right)(t)
	+\bm{S}_g^q(t) \cdot\left(\alpha^R * \bm{S}_g^c \right)(t) \right],
\end{align}
where $\bm{S}_g^c(t)$ and $\bm{S}_g^q(t)$ include only up to first-order corrections in quantum components.
Assuming that $\alpha^{A/R}$ can be written in terms of distributions, we can define the distribution $\adis(t)= -\alpha^R(t) -\alpha^A(-t)$ and rewrite the semi-classical action as
\begin{align}
	iS_{sc}\left[g,\bm{\xi}\right] &= i \int dt \; S\Big[   -\theta_q\sin\theta_c (-B_z+KS\cos\theta_c +\dot{\phi}_c ) -\theta_q\cos\theta_c(B_x\cos\phi_c +B_y\sin\phi_c) \nonumber\\
	&+\phi_q\sin\theta_c(\dot{\theta}_c +B_x\sin\phi_c- B_y\cos\phi_c )	 \Big] + i \int dt\;\left[  \left(\adis * \bm{S}_g^c \right)(t) +\bm{\xi}(t) 	\right]\bm{S}_g^q(t).
\end{align}
Recall that, using the Euler angles, we have $\int Dg= \int D\theta  D\phi \sin(\theta)$. Technically, the factor of $\sin(\theta)$ would end up in the action. However, since one could define $\rho = \cos(\theta)$ as a new variable in order to avoid this, we know that this term is not relevant to the physics. Hence, we can disregard it. 

Since all terms in $iS_{sc}\left[g,\bm{\xi}\right]$ are either linear in $\theta_q$ or $\phi_q$, we find two Euler-Lagrange equations of the form
\begin{equation}
	\frac{\delta \mathcal{L}_{sc}}{\delta \theta_q} =0 \quad \text{and}\quad \frac{\delta \mathcal{L}_{sc}}{\delta \phi_q} =0.
\end{equation}
Remembering Eq. \eqref{eq:qcrelation}, we see that $\frac{\delta \bm{S}_g^q(t)}{\delta \theta_q} = \frac{\delta \bm{S}_g^c(t)}{\delta \theta_c} $ and $\frac{\delta \bm{S}_g^q(t)}{\delta \phi_q} = \frac{\delta \bm{S}_g^c(t)}{\delta \phi_c} $. Hence, the e.o.m. can be rearranged to yield
\begin{align}
	\dot{\phi}_c &=  \frac{1}{S\sin\theta_c}
	\left[ -\bm{B}(S_z^c) +\left(\adis * \bm{S}_g^c \right)(t) +\bm{\xi}(t) 	\right]\cdot\frac{\delta \bm{S}_g^c(t)}{\delta \theta_c}  \label{eq:eomphi1}
\end{align}
and
\begin{align}
	\dot{\theta}_c   &=  -\frac{1}{S\sin\theta_c}
	\left[ -\bm{B}(S_z^c) +\left(\adis * \bm{S}_g^c \right)(t) +\bm{\xi}(t) 	\right]\cdot\frac{\delta \bm{S}_g^c(t)}{\delta \phi_c}, \label{eq:eomtheta1}
\end{align}
where $\bm{B}(S_z^c)= \begin{pmatrix}
	B_x \\ B_y \\ B_z -K S_z^c
\end{pmatrix}$.

\subsection{Generalized Landau-Lifshitz-Gilbert equation}
\label{sec:sphericalllg}
We want to show that the equations found by the microscopic model are in fact precisely of the LLG form. For this, we will have to start from the LLG equation, and introduce the same two Euler angles $\theta$ and $\phi$ for the spin, and show that this gives rise to the same set of equations as previously deduced. 

We begin with the generalized LLG equation 
\begin{equation}
	\dot{\bm{S}}(t)= \bm{S}(t) \times \left[-\bm{B}(S_z)+ \left(\adis * \bm{S} \right)(t) +\bm{\xi}(t) \right], \label{eq:gllg}
\end{equation}
where $\adis(t)= -\alpha^R(t) -\alpha^A(-t)$, $	\langle \bm{\xi}(t) \bm{\xi}(t') \rangle = -2i \alpha^K(t-t')$ and $\bm{B}(S_z)= (B_x , B_y , B_z -K S_z )^T$.
Since the velocity of $\bm{S}$ is always perpendicular to $\bm{S}$, we know that the magnitude of $\bm{S}$ is constant. Hence, we can go to spherical coordinates, such that
\begin{equation}
	\bm{S} = S \begin{pmatrix}
		\sin\theta \cos\phi \\ \sin\theta \sin\phi \\ \cos\theta
	\end{pmatrix}.
\end{equation}
Inserting this into the LLG equation, we firstly see that
\begin{align*}
	\dot{\bm{S}} &=\dot{\theta}\frac{\partial \bm{S}}{\partial \theta} +\dot{\phi} \frac{\partial \bm{S}}{\partial \phi} =  
	\dot{\theta}S \begin{pmatrix}
		\cos\theta \cos\phi \\ \cos\theta \sin\phi \\ -\sin\theta
	\end{pmatrix}
	+\dot{\phi}S \begin{pmatrix}
		-\sin\theta \sin\phi \\ \sin\theta \cos\phi \\ 0
	\end{pmatrix}.
\end{align*}
Now, we notice that the RHS of the LLG equation can, without loss of generality, be written as $\bm{S}(t) \times \bm{r}$ with $ \bm{r}=(x,y,z)^T$. Working this out explicitly, we find that the LLG equation $\dot{\bm{S}}=\bm{S} \times \bm{r}$ becomes
\begin{equation}
	S \begin{pmatrix}
		\dot{\theta}\cos\theta \cos\phi -\dot{\phi}\sin\theta \sin\phi \\ \dot{\theta}\cos\theta \sin\phi+\dot{\phi}\sin\theta \cos\phi \\ -\dot{\theta}\sin\theta
	\end{pmatrix}
	=
	S\begin{pmatrix}
		z\sin\theta \sin\phi - y\cos\theta \\
		x\cos\theta - z \sin\theta \cos\phi \\
		y\sin\theta \cos\phi -x\sin\theta \sin\phi
	\end{pmatrix}.
\end{equation}
We note that the equation corresponding to the $z$ component can be written as
\begin{equation}
	\dot{\theta}= -\frac{1}{\sin\theta} \bm{r} \cdot \begin{pmatrix}
		-\sin\theta \sin\phi \\ \sin\theta \cos\phi \\ 0
	\end{pmatrix}
	=
	-\frac{1}{S\sin\theta} \bm{r} \cdot\frac{\partial \bm{S}}{\partial \phi}.
\end{equation}
Now, we add up the $\hat{x}$ and $\hat{y}$ equations, such that the $\dot{\theta}$ cancels (i.e. $-\hat{x}\sin\phi+\hat{y}\cos\phi$). This yields
\begin{align}
	\dot{\phi}\sin\theta (\sin^2\phi + 	
	\cos^2\phi)&= 
	-z\sin\theta (\sin^2\phi+\cos^2\phi) + y\cos\theta\sin\phi +
	x\cos\theta\cos\phi,  
	\nonumber\\\intertext{which simplifies to}
	\dot{\phi} &= \frac{1}{\sin\theta}
	\bm{r} \cdot\begin{pmatrix}
		\cos\theta \cos\phi \\ \cos\theta \sin\phi \\ -\sin\theta
	\end{pmatrix} =\frac{1}{S\sin\theta}
	\bm{r} \cdot \frac{\partial \bm{S}}{\partial \theta}.
\end{align}
By inserting $\bm{r}=-\bm{B}(S_z)+ \left(\adis * \bm{S} \right)(t) +\bm{\xi}(t)$, we see that this is identical to the equations derived from the microscopic model
\begin{align}
	\dot{\phi}_c &=  \frac{1}{S\sin\theta_c}
	\left[-\bm{B}(S_z) + \left(\adis * \bm{S}_g^c \right)(t) +\bm{\xi}(t) 	\right]\cdot\frac{\delta \bm{S}_g^c(t)}{\delta \theta_c}; \label{eq:eomphi2}
	\\
	\dot{\theta}_c   &= - \frac{1}{S\sin\theta_c}
	\left[-\bm{B}(S_z) + \left(\adis * \bm{S}_g^c \right)(t) +\bm{\xi}(t) 	\right]\cdot\frac{\delta \bm{S}_g^c(t)}{\delta \phi_c}. \label{eq:eomtheta2}
\end{align}
Therefore, we may conclude that our microscopic model is described by the generalized LLG equation.

	For the fractional LLG equation, we are in particular interested in the case where  $\adis * \bm{S} = \alpha_s D_t^s \bm{S}$, where $D_t^s$ is a fractional derivative. For instance, assuming $0<s<1$, the Liouville fractional derivative is given by 
\begin{equation}
	D_t^s f(t)= \frac{1}{\Gamma(1-s)} \int_{-\infty}^t (t-t')^{-s} f'(t') dt'.
\end{equation}
So, if $\adis = \frac{\alpha_s\Theta(t)}{\Gamma(1-s)} t^{-s} \partial_t  $, then, because of the convolution with $\bm{S}$, we would find a fractional LLG equation, whereas $\adis = \alpha_1 \delta(t) \partial_t $ would give the  regular LLG equation.

\section{Fractional derivative from non-Ohmic spectral function}

Here, we will compute the type of dissipation which comes from the spectral function. We will first derive the spectral function from microscopic quantities to see how it ends up in the Greens function. Then, we will calculate the dissipation for three different cases.

	\subsection{Calculating the effective Greens functions}
We recall that 
\begin{equation*}
	\alpha^{A/R}(t-t') =   \sum_\alpha \left( \frac{\gamma_\alpha^2}{4} G_\alpha^{A/R}(t-t') +\delta(t-t')\frac{ \gamma_\alpha^2}{2m_\alpha \omega_\alpha^2}\right),
\end{equation*} 
where 
\begin{equation*}
	[G^{-1}_\alpha]^{R/A}(t-t') = \delta(t-t') \frac{m_\alpha}{2}[(i\partial_t \pm i0)^2 -\omega_\alpha^2].
\end{equation*} 
By the fluctuation dissipation theorem, we also have 
\begin{equation*}
	\alpha^K(\omega)=\left[\alpha^R(\omega) -\alpha^A(\omega)\right]\coth\left(\frac{\omega}{2T}\right).
\end{equation*}
We are interested in finding closed forms for $\alpha^{R/A/K}(t-t')$.
Using the relation
\begin{equation}
	\int dt'G^{-1}(t-t') G(t'-t'')=\delta(t-t''),
\end{equation}
we note that 
\begin{equation}
	\frac{m_\alpha}{2}[(i\partial_t \pm i0)^2 -\omega_\alpha^2]G^{R/A}_\alpha(t-t'') = \delta(t-t'').
\end{equation}
The Fourier transform\footnote{We use the convention $f(\omega)=\int_{-\infty}^\infty dt\;e^{i\omega t} f(t)$ and $f(t)=\frac{1}{2\pi}\int_{-\infty}^\infty d\omega\; e^{-i\omega t} f(\omega) $, with $\delta(t)= \frac{1}{2\pi}\int_{-\infty}^\infty d\omega\; e^{i\omega t}$.} yields
\begin{equation}
	G^{R/A}_\alpha(\omega) =  \frac{1}{\frac{m_\alpha}{2}\left[(\omega \pm i0)^2 -\omega_\alpha^2\right]}.
\end{equation}
We therefore find that 
\begin{align}
	\alpha^{R/A}(\omega)&= 
	 \sum_\alpha \left( \frac{\gamma_\alpha^2}{4} G_\alpha^{A/R}(\omega) +\frac{ \gamma_\alpha^2}{2m_\alpha \omega_\alpha^2}\right) 
	=\sum_\alpha \frac{\gamma_\alpha^2}{2m_\alpha}\left[ \frac{1}{(\omega \pm i0)^2 -\omega_\alpha^2}+\frac{1}{\omega_\alpha^2}\right]
	=\sum_\alpha \frac{\gamma_\alpha^2}{2m_\alpha\omega_\alpha^2} \frac{\omega^2}{(\omega \pm i0)^2 -\omega_\alpha^2}.
\end{align} 
The spectral function is given by the imaginary part of the Fourier transform of the dynamical susceptibility 
\begin{equation}
	\chi(\omega)= \frac{\delta}{\delta \bm{S}(\omega)} \sum_\alpha \gamma_\alpha \hat{\bm{x}}_\alpha(\omega) .
\end{equation}
The Hamiltonian e.o.m. for the bath can be found by going back to our starting Hamiltonian
\begin{align*}
	H &= \bm{B}\cdot\bm{S} +\sum_\alpha \gamma_\alpha \bm{S} \cdot\hat{\bm{x}}_\alpha + \sum_\alpha \frac{\hat{\bm{p}}_\alpha^2}{2m_\alpha} +\frac{m_\alpha \omega_\alpha^2}{2} \hat{\bm{x}}_\alpha^2+ \sum_\alpha \frac{ \gamma_\alpha^2}{2m_\alpha \omega_\alpha^2} \bm{S}^2.
\end{align*}
The e.o.m. reads
\begin{equation}
	\dot{\hat{\bm{x}}}_\alpha=\frac{\hat{\bm{p}}_\alpha}{m_\alpha } \quad\text{and} \quad 	\dot{\hat{\bm{p}}}_\alpha=-\gamma_\alpha \bm{S} -m_\alpha \omega_\alpha^2 \hat{\bm{x}}_\alpha.
\end{equation}
Combining both equations and taking the Fourier transform, we find
\begin{equation}
	\hat{\bm{x}}_\alpha(\omega)= \frac{\gamma_\alpha}{m_\alpha [(\omega+i0)^2-\omega_\alpha^2]} \bm{S}(\omega),
\end{equation}
where we have taken an infinitesimal amount of dissipation $+i0$ on the oscillators into account. This leads to
\begin{equation}
	\chi(\omega)=  \sum_\alpha \frac{\gamma_\alpha^2}{m_\alpha [(\omega+i0)^2-\omega_\alpha^2]} =  \sum_\alpha \frac{\gamma_\alpha^2}{2 m_\alpha \omega_\alpha} \left( \frac{1}{\omega+i0-\omega_\alpha}-\frac{1}{\omega+i0+\omega_\alpha} \right).
\end{equation}
We remark that 
\begin{equation}
	\operatorname{Im} \frac{1}{x+i0}= -\pi\delta(x),
\end{equation}
which leads to
\begin{equation}
	J(\omega)= \operatorname{Im}\chi(\omega)= -\sum_\alpha \frac{\pi\gamma_\alpha^2}{2 m_\alpha \omega_\alpha} \left[ \delta(\omega-\omega_\alpha)-\delta(\omega+\omega_\alpha) \right]= -\frac{\pi}{2}\sum_\alpha \frac{\gamma_\alpha^2}{ m_\alpha \omega_\alpha}  \delta(\omega-\omega_\alpha),
\end{equation}
where we used that all oscillator frequencies are positive.
We can identify the spectral function in $\alpha^{R/A}$ as
\begin{equation}
	\alpha^{R/A}(\omega)=  \sum_\alpha \frac{\gamma_\alpha^2}{2m_\alpha\omega_\alpha^2} \frac{\omega^2}{(\omega \pm i0)^2 -\omega_\alpha^2}
	= -\int_0^\infty \frac{d\varepsilon}{\pi} \frac{ \omega^2 \varepsilon^{-1} J(\varepsilon) }{(\omega \pm i0)^2 -\varepsilon^2}.
\end{equation}
Now, we will assume a particular shape for $J(\varepsilon)$. This can be either Ohmic or non-Ohmic, but in general we may assume a power-law behavior as some $ J(\varepsilon) = \alpha_s \varepsilon^s$.

\subsection{Ohmic spectral function}
Beginning with the Ohmic case $ J(\varepsilon) = \alpha_1 \varepsilon$, we see that
\begin{align}
	2i \text{ Im } \alpha^{R/A}(\omega) 	&= 
	\alpha^{R/A}(\omega)-\left[\alpha^{R/A}\right]^*(\omega)\nonumber\\
	&=
	-\alpha_1\int_0^\infty \frac{d\varepsilon}{\pi}\left[ \frac{ \omega^2  }{(\omega \pm i0)^2 -\varepsilon^2} -\frac{ \omega^2  }{(\omega \mp i0)^2 -\varepsilon^2}\right]\nonumber\\
	&= 
	-\alpha_1\int_{-\infty}^\infty \frac{d\varepsilon}{2\pi} \left[\frac{ \omega^2  }{(\omega \pm i0)^2 -\varepsilon^2} -\frac{ \omega^2  }{(\omega \mp i0)^2 -\varepsilon^2}\right]\nonumber\\
	&=
	-\alpha_1\omega^2\int_{-\infty}^\infty \frac{d\varepsilon}{2\pi}  \left[\frac{ 1  }{(\omega \pm i0+\varepsilon)(\omega \pm i0-\varepsilon)} -\frac{ 1  }{(\omega \mp i0+\varepsilon)(\omega \mp i0-\varepsilon)}\right]\nonumber\\
	&=
	\pm 4i0\omega^3 \alpha_1\int_{-\infty}^\infty \frac{d\varepsilon}{2\pi}  \frac{ 1 }{(\omega \pm i0+\varepsilon)(\omega \pm i0-\varepsilon)(\omega \mp i0+\varepsilon)(\omega \mp i0-\varepsilon)}, 
\end{align}
which has four poles at $\varepsilon=\pm_1 (\omega \pm_2 i0)$. Since the integral scales as $<1/|\varepsilon|$, we can add an infinite radius half circle to complete a complex contour integral. Notice from symmetry that we will always have one of each of the four poles. We can thus drop the $\pm$ signs inside since this only changes the notation order in the fraction. We thus find that
\begin{align}
	2i \text{ Im } \alpha^{R/A}(\omega) 
	&=
	\pm 4i0\omega^3 \alpha_1\int_{-\infty}^\infty \frac{d\varepsilon}{2\pi}  \frac{ 1 }{(\varepsilon+\omega + i0)(\varepsilon-\omega - i0)(\varepsilon+ \omega - i0)(\varepsilon-\omega + i0)}.
\end{align}
Completing the contour along the top, we find poles at $\varepsilon=\pm \omega +i0$, which yields
\begin{align}
	2i \text{ Im } \alpha^{R/A}(\omega) 
	=
	\mp0\cdot4 \omega^3 \alpha_1& \left[
	\frac{1 }{ (\omega +i0+\omega + i0)(\omega +i0+ \omega - i0)(\omega +i0-\omega + i0)} \right.\nonumber\\
	&+\left.
	\frac{ 1 }{(-\omega +i0+\omega + i0)(-\omega +i0-\omega - i0)(-\omega +i0-\omega + i0)}
	\right]\nonumber\\
	&=
	\mp 0\cdot4 \omega^3 \alpha_1 \left[
	\frac{ 1 }{2(\omega +i0)(2\omega)(2i0)} +
	\frac{ 1 }{(2i0)(-2\omega)2(-\omega +i0)}
	\right]\nonumber\\
	&=
	\pm \frac{i}{2} \omega^3 \alpha_1 \left[
	\frac{ 1 }{\omega^2 +i0\omega} +
	\frac{ 1 }{\omega^2- i0\omega}
	\right]\nonumber\\
	&=
	\pm \frac{i}{2} \omega^3 \alpha_1 \left[
	\frac{ \omega^2- i0\omega + \omega^2 +i0\omega}{\omega^4 } 
	\right]\nonumber\\
	&=\pm i\omega \alpha_1.
\end{align}
Hence, $\text{ Im } \alpha^{R/A}(\omega) =\pm \alpha_1 \omega /2$.
Similarly,
\begin{align}
	2 \text{ Re } \alpha^{R/A}(\omega) 	&= 
	\alpha^{R/A}(\omega)+\left[\alpha^{R/A}\right]^*(\omega)\nonumber\\
	&=
	-\alpha_1\int_0^\infty \frac{d\varepsilon}{\pi}\left[ \frac{ \omega^2  }{(\omega \pm i0)^2 -\varepsilon^2} +\frac{ \omega^2  }{(\omega \mp i0)^2 -\varepsilon^2}\right]\nonumber\\
	&= 
	-\alpha_1\int_{-\infty}^\infty \frac{d\varepsilon}{2\pi} \left[\frac{ \omega^2  }{(\omega \pm i0)^2 -\varepsilon^2} +\frac{ \omega^2  }{(\omega \mp i0)^2 -\varepsilon^2}\right]\nonumber\\
	&=
	-\alpha_1\int_{-\infty}^\infty \frac{d\varepsilon}{2\pi}  \left[\frac{ \omega^2 }{(\omega \pm i0+\varepsilon)(\omega \pm i0-\varepsilon)} +\frac{ \omega^2 }{(\omega \mp i0+\varepsilon)(\omega \mp i0-\varepsilon)}\right]\nonumber\\
	&=
	\alpha_1\int_{-\infty}^\infty \frac{d\varepsilon}{\pi}  \frac{ \omega^2 (\varepsilon^2-\omega^2) }{(\varepsilon+\omega + i0)(\varepsilon-\omega - i0)(\varepsilon+ \omega - i0)(\varepsilon-\omega + i0)}.
\end{align}
Since the integral scales as $1/|\varepsilon|$, we can  freely add the infinite circular contour along the top. Applying the residue theorem, we find
\begin{align}
	2 \text{ Re } \alpha^{R/A}(\omega) 	&= 2i\alpha_1   \left[
	\frac{ \omega^2 ((\omega+i0)^2-\omega^2) }{(\omega+i0+\omega + i0)(\omega+i0+ \omega - i0)(\omega+i0-\omega + i0)} \right. \nonumber\\
	&\quad \left.+
	\frac{ \omega^2 ((-\omega+i0)^2-\omega^2) }{(-\omega+i0+\omega + i0)(-\omega+i0-\omega - i0)(-\omega+i0-\omega + i0)} \right]\nonumber\\
	&=
	2\alpha_1  i \left[
	\frac{ \omega^2 (2\omega i0) }{2(\omega+i0)(2\omega)(2i0)} 
	+
	\frac{ \omega^2 (-2\omega i0) }{(2i0)(-2\omega)2(-\omega+i0)} \right]\nonumber\\
	&=
	\frac{\alpha_1}{2}  i \left[
	\frac{ \omega^2  }{\omega+i0} 
	-
	\frac{ \omega^2  }{\omega-i0} \right]\nonumber\\
	&=
	\frac{\alpha_1}{2}  i \left[
	\frac{ \omega^2(-2i0)  }{\omega^2} 
	\right]\nonumber\\
	&= \alpha_1 0,
\end{align}
which means that $\text{ Re } \alpha^{R/A}(\omega) = \frac{\alpha_1}{2} \cdot 0=0$. Hence,
\begin{equation}
	\alpha^{R/A}(\omega) = \pm \frac{\alpha_1 i\omega}{2}
\end{equation}
and since $\adis(t)=-\alpha^R(t) -\alpha^A(-t)$ we have
\begin{equation}
	\adis(\omega)= -\alpha^R(\omega) -\alpha^A(-\omega) =-\alpha_1 i\omega.
\end{equation}
In the LLG equation, we thus find
\begin{align}
	\left(\adis * \bm{S} \right)(t)&= 
	\frac{1}{2\pi} \int d\omega e^{-i\omega t}\left(\adis * \bm{S} \right)(\omega) \nonumber\\
	&= 
	\frac{1}{2\pi} \int d\omega e^{-i\omega t}\adis(\omega) \bm{S}(\omega) \nonumber\\
	&= 
	\frac{1}{2\pi} \int d\omega e^{-i\omega t}(-\alpha_1 i \omega )\bm{S}(\omega) \nonumber\\
	&= 
	\alpha_1 \frac{\partial}{\partial t}\frac{1}{2\pi} \int d\omega e^{-i\omega t} \bm{S}(\omega) \nonumber\\
	&= \alpha_1 \dot{\bm{S}}(t)\label{ohmicdiss}.
\end{align}
Furthermore, we have that 
\begin{align}
	\alpha^K(\omega) &= [\alpha^R(\omega) -\alpha^A(\omega)]\coth\left( \frac{\omega}{2T} \right)\nonumber\\
	&= i\alpha_1 \omega \coth\left( \frac{\omega}{2T} \right).
\end{align}
We have two regimes from the cotangent which crossover around $\omega \approx 2T$. If the temperature is large enough that we can approximate $\coth\left( \frac{\omega}{2T} \right) \approx \frac{2T}{\omega}$, then we have $\alpha^K(\omega) \approx 2i\alpha_1 T$. Therefore, we find 
\begin{equation}
	\alpha^K(t)= 2i\alpha_1  T \frac{1}{2\pi} \int_{-\infty}^{\infty} d\omega\; e^{-i\omega t} = 2i\alpha_1 T \delta(t),
\end{equation}
and thus
\begin{equation}
	\langle \bm{\xi}_m(t) \bm{\xi}_n(t') \rangle = -2i \delta_{m,n}\alpha^K(t-t') = 4\alpha_1 T \delta_{m,n} \delta(t-t').\label{ohmicnoise}
\end{equation}	
We see that Eq.~\eqref{ohmicdiss} and Eq.~\eqref{ohmicnoise} combine to give the regular LLG equation with first-order dissipation and white noise fluctuation:
\begin{equation}
	\dot{\bm{S}}(t) = \bm{S}(t) \times \left[ -\bm{B} +\alpha_1\dot{\bm{S}}(t) +\bm{\xi}(t) \right] .
\end{equation}

\subsection{Sub-Ohmic spectral function}
We now consider the case where 
\begin{equation}
	J(\varepsilon)=\alpha_s\sin\left(\frac{\pi s}{2}\right) \varepsilon^s,
\end{equation} with $0<s<1$. In this case, 
\begin{equation}
	\alpha^{R/A}(\omega)=   -\alpha_s\sin\left(\frac{\pi s}{2}\right)\int_0^\infty \frac{d\varepsilon}{\pi} \frac{ \omega^2\varepsilon^{s-1}  }{(\omega \pm i0)^2-\varepsilon^2 }.
\end{equation}
Considering the LLG equation, the relevant dissipation term is $(\adis * \bm{S})(t)$. In terms of the Fourier transform of $\adis(t)$, we find that
\begin{align}
	(\adis * \bm{S})(t) &= \int_{-\infty}^{\infty} dt' \adis(t-t') \bm{S}(t') \nonumber\\
	&=
	-\int_{-\infty}^{\infty} dt' [\alpha^R(t-t') + \alpha^A(t'-t) ]\bm{S}(t') \nonumber\\
	&=
	-\frac{1}{2\pi}\int_{-\infty}^{\infty} dt' \int_{-\infty}^\infty d\omega \left[e^{-i\omega (t-t')} \alpha^R(\omega) +e^{i\omega (t-t')} \alpha^A(\omega) \right] \bm{S}(t') \nonumber\\
	&=
	-\frac{1}{2\pi}\int_{-\infty}^{\infty} dt' \int_{-\infty}^\infty d\omega \;  e^{-i\omega (t-t')} \left[\alpha^R(\omega) + \alpha^A(-\omega) \right] \bm{S}(t')\nonumber\\
	&=
	\alpha_s\frac{\sin\left(\frac{\pi s}{2}\right)}{2\pi^2}\int_{-\infty}^{\infty} dt' 
	\int_{-\infty}^\infty d\omega \int_0^\infty d\varepsilon\; e^{-i\omega (t-t')} \left[ \frac{ \omega^2\varepsilon^{s-1}  }{(\omega + i0)^2-\varepsilon^2}+\frac{ (-\omega)^2\varepsilon^{s-1}  }{(-\omega - i0)^2-\varepsilon^2}
	\right] \bm{S}(t')\nonumber\\
	&=
	-\alpha_s\frac{\sin\left(\frac{\pi s}{2}\right)}{\pi^2}\int_{-\infty}^{\infty} dt' 
	\int_0^\infty d\varepsilon \int_{-\infty}^\infty d\omega \left[e^{-i\omega (t-t')}  \frac{ \omega^2\varepsilon^{s-1} }{\varepsilon^2 -(\omega + i0)^2}\bm{S}(t')\right].
\end{align}
Notice that we have two poles at $\omega=\pm\varepsilon -i0$, below the real axis. If $t-t'<0$, then the exponential will go to zero as $\omega \to +i\infty$. Hence we could close the $\omega$ integration with a complex contour as an infinite half-circle along the top, and get zero from the Cauchy theorem. If $t-t'>0$, however, we see that the exponential goes to zero when $\omega\to -i\infty$. Hence, we can close the $\omega$ integration along the bottom. Thus, using the residue theorem (reversing the integration direction), we find
\begin{align}
	(\adis * \bm{S})(t) &=
	-\alpha_s\frac{\sin\left(\frac{\pi s}{2}\right)}{\pi^2}\int_{-\infty}^{\infty} dt' 
	\int_0^\infty d\varepsilon \;2\pi i \Theta(t-t')\nonumber\\
	&\qquad\Big[
	e^{-i(\varepsilon -i0) (t-t')}  \frac{ (\varepsilon -i0)^2\varepsilon^{s-1}  }{(\varepsilon -i0+i0 +\varepsilon)}
	+
	e^{-i(-\varepsilon -i0) (t-t')}  \frac{ (-\varepsilon -i0)^2\varepsilon^{s-1}  }{(-\varepsilon -i0 +i0 -\varepsilon)}
	\Big]\bm{S}(t')\nonumber\\
	&=
	-i\alpha_s\frac{\sin\left(\frac{\pi s}{2}\right)}{\pi}\int_{-\infty}^{t} dt' 
	\int_0^\infty d\varepsilon \left[
	e^{-i\varepsilon  (t-t')}  \frac{ \varepsilon^{s+1}  }{\varepsilon }
	+
	e^{i\varepsilon  (t-t')}  \frac{\varepsilon^{s+1}  }{-\varepsilon}
	\right]\bm{S}(t')\nonumber\\
	&=
	-i\alpha_s\frac{\sin\left(\frac{\pi s}{2}\right)}{\pi} \int_{-\infty}^{t} dt' 
	\int_0^\infty d\varepsilon \left[
	e^{-i\varepsilon  (t-t')}  
	-
	e^{i\varepsilon  (t-t')} 
	\right] \varepsilon^{s}\bm{S}(t')\nonumber\\ 
	&= 
	-2\alpha_s\frac{\sin\left(\frac{\pi s}{2}\right)}{\pi} \int_{-\infty}^{t} dt' 
	\int_0^\infty d\varepsilon \sin[\varepsilon(t-t')] \varepsilon^{s}\bm{S}(t')\nonumber\\ 
	&= 
	-2\alpha_s\frac{\sin\left(\frac{\pi s}{2}\right)}{\pi}  
	\int_0^\infty d\varepsilon \left\{
	\left[\varepsilon^{s-1}\cos[\varepsilon(t-t')] \bm{S}(t')\right]_{t'=t_0}^{t'=t}
	-\int_{t_0}^{t} dt' \cos[\varepsilon(t-t')] \varepsilon^{s-1}\dot{\bm{S}}(t')\right\}\nonumber\\ 
	&= 
	-2\alpha_s\frac{\sin\left(\frac{\pi s}{2}\right)}{\pi}  
	\int_0^\infty d\varepsilon \left\{
	\varepsilon^{s-1}\bm{S}(t) -\varepsilon^{s-1}\cos[\varepsilon (t-t_0)] \bm{S}(t_0)
	-\int_{t_0}^{t} dt' \cos[\varepsilon(t-t')] \varepsilon^{s-1}\dot{\bm{S}}(t')\right\}. \label{eq:subohmicmiddle}
\end{align}
The first term vanishes because of the cross product with $\bm{S}(t)$ in the LLG equation. The second term is where we had to be careful. Here, we should realize that the $-\infty$ is physically only indicating that it is a time very far in the past. So, to avoid unphysical infinities, we introduced a finite initial time $t_0$ and we will take $t_0\to-\infty$ later. For this, we need to introduce some fractional derivative notation. We define the Riemann-Liouville~(RL) and Caputo~(C) derivatives of order $s$, with an integer $n$ such that $n\le s<n+1$, as
\begin{align}
	\prescript{RL}{t_0}{D}_t^s f(t) &= \frac{d^n}{dt^n} \frac{1}{\Gamma(n-s)} \int_{t_0}^t dt' (t-t')^{n-1-s} f(t'),\\
	\prescript{C}{t_0}{D}_t^s f(t) &=  \frac{1}{\Gamma(n-s)} \int_{t_0}^t dt' (t-t')^{n-1-s} f^{(n)}(t'),
\end{align}
where we reserve the simpler $D_t^s$ notation for the Liouville derivative that was used in the main text.

Now, rescaling $\varepsilon\to \varepsilon/(t-t_0)$ and $\varepsilon\to \varepsilon/(t-t')$ in the second and third terms of Eq.~\eqref{eq:subohmicmiddle} respectively, we have
\begin{align}
	(\adis * \bm{S})(t) 
	&= 
	2\alpha_s\frac{\sin\left(\frac{\pi s}{2}\right)}{\pi}  
	\int_0^\infty d\varepsilon \cos(\varepsilon) \varepsilon^{s-1}\left[(t-t_0)^{-s}\bm{S}(t_0)
	+\int_{t_0}^t dt' (t-t')^{-s} \dot{\bm{S}}(t')\right]\nonumber \\
	&= 
	2\alpha_s\frac{\sin\left(\frac{\pi s}{2}\right)}{\pi}  
	\left[\cos\left(\frac{\pi s}{2}\right) \Gamma(s) \right] \left[(t-t_0)^{-s}\bm{S}(t_0)
	+\int_{t_0}^t dt' (t-t')^{-s} \dot{\bm{S}}(t')\right]\nonumber \\
	&= 
	2\alpha_s\frac{\sin\left(\frac{\pi s}{2}\right)}{\pi}  
	\cos\left(\frac{\pi s}{2}\right) \Gamma(s)  \left[(t-t_0)^{-s}\bm{S}(t_0)
	+\Gamma(1-s) \prescript{C}{t_0}{D}_t^s \bm{S}(t)\right]\nonumber \\
	&= 
	2\alpha_s\frac{\sin\left(\frac{\pi s}{2}\right)}{\pi}  
	\cos\left(\frac{\pi s}{2}\right) \Gamma(s) \Gamma(1-s) \left[\frac{(t-t_0)^{-s}}{\Gamma(1-s)}\bm{S}(t_0)
	+ \prescript{C}{t_0}{D}_t^s \bm{S}(t)\right]\nonumber \\
	&= 
	2\alpha_s\frac{\sin\left(\frac{\pi s}{2}\right)}{\pi}  
	\cos\left(\frac{\pi s}{2}\right) \frac{\pi}{\sin(\pi s)} \prescript{RL}{t_0}{D}_t^s \bm{S}(t)\nonumber \\
	&= 
	\alpha_s \prescript{RL}{t_0}{D}_t^s \bm{S}(t) 
	= 
	\alpha_s D_t^s \bm{S}(t),
\end{align}
where we used several identities from Sec. 6 in the Sup. Mat. of Ref.~\cite{Verstraten2021} and in the last line we sent $t_0\to-\infty$. 

For the noise correlation, we have to compute the Keldysh component. This is
\begin{align}
	\alpha^K(t) &= \frac{1}{2\pi}\int_{-\infty}^\infty d\omega\; e^{-i\omega t} \alpha^K(\omega) \nonumber\\
	&=
	\frac{1}{2\pi}\int_{-\infty}^\infty d\omega\; e^{-i\omega t}[\alpha^R(\omega) -\alpha^A(\omega)]\coth\left( \frac{\omega}{2T} \right)\nonumber\\
	&=
	\alpha_s\frac{\sin\left(\frac{\pi s}{2}\right)}{2\pi^2}\int_0^\infty d\varepsilon\int_{-\infty}^\infty d\omega\; e^{-i\omega t} \left[\frac{ \omega^2\varepsilon^{s-1}  }{\varepsilon^2 -(\omega + i0)^2} -\frac{ \omega^2\varepsilon^{s-1}  }{\varepsilon^2 -(\omega - i0)^2}\right]\coth\left( \frac{\omega}{2T} \right)\nonumber\\
	\intertext{(now send $\omega\to -\omega$ in the advanced part)}
	&=
	\alpha_s\frac{\sin\left(\frac{\pi s}{2}\right)}{2\pi^2}\int_0^\infty d\varepsilon\int_{-\infty}^\infty d\omega\;  \left[\frac{e^{-i\omega t} \omega^2\varepsilon^{s-1}  }{\varepsilon^2 -(\omega + i0)^2} +\frac{e^{i\omega t} \omega^2\varepsilon^{s-1}  }{\varepsilon^2 -(-\omega - i0)^2}\right]\coth\left( \frac{\omega}{2T} \right)\nonumber\\
	&=
	\alpha_s\frac{\sin\left(\frac{\pi s}{2}\right)}{\pi^2}\int_0^\infty d\varepsilon\int_{-\infty}^\infty d\omega\;  \cos(\omega t)\frac{ \omega^2\varepsilon^{s-1}  }{\varepsilon^2 -(\omega + i0)^2}\coth\left( \frac{\omega}{2T} \right).
\end{align}
Now, we send $\omega \to \omega/t$ and $\varepsilon\to \varepsilon/t$, which yields
\begin{align}
	\alpha^K(t) &=
	\alpha_s t^{-1-s}\frac{\sin\left(\frac{\pi s}{2}\right)}{\pi^2}\int_0^\infty d\varepsilon\int_{-\infty}^\infty d\omega\;  \cos(\omega )\frac{ \omega^2\varepsilon^{s-1}  }{\varepsilon^2 -(\omega + i0)^2}\coth\left( \frac{\omega}{2tT} \right).
\end{align}
Taking the high temperature limit, we get
\begin{align}
	\alpha^K(t) &=
	\alpha_s t^{-1-s}\frac{\sin\left(\frac{\pi s}{2}\right)}{\pi^2}\int_0^\infty d\varepsilon\int_{-\infty}^\infty d\omega\;  \cos(\omega )\frac{ \omega^2\varepsilon^{s-1}  }{\varepsilon^2 -(\omega + i0)^2} \frac{2tT}{\omega} \nonumber\\
	&=
	2T \alpha_s t^{-s}\frac{\sin\left(\frac{\pi s}{2}\right)}{\pi^2}\int_0^\infty d\varepsilon\int_{-\infty}^\infty d\omega\;  \cos(\omega )\frac{ \omega\varepsilon^{s-1}  }{\varepsilon^2 -(\omega + i0)^2} \nonumber\\
	&=
	-2T \alpha_s t^{-s}\frac{\sin\left(\frac{\pi s}{2}\right)}{\pi^2}\int_0^\infty d\varepsilon\int_{-\infty}^\infty d\omega\;  \frac{ \cos(\omega )\omega\varepsilon^{s-1}  }{(\omega +i0 -\varepsilon)(\omega+i0 +\varepsilon)}.  
\end{align}
Now, we want to close the integral over $\omega$ with an infinite half-circle. For this, we need fast enough convergence of the integrand to zero. Splitting the cosine into two exponential parts $\cos(\omega)= (e^{i\omega} +e^{-i\omega})/2$, we see that the first term goes to zero when $\omega\to i\infty$ and that the second term goes to zero when $\omega\to-i\infty$. Since we have poles at $\omega=\pm\varepsilon -i0$, the integral along the top half-plane vanishes. The integral along the bottom is then computed as
\begin{align}
	\alpha^K(t) &=
	-T \alpha_s t^{-s}\frac{\sin\left(\frac{\pi s}{2}\right)}{\pi^2}\int_0^\infty d\varepsilon\int_{-\infty}^\infty d\omega\;  \frac{ e^{-i\omega} \omega\varepsilon^{s-1}  }{(\omega +i0 -\varepsilon)(\omega+i0 +\varepsilon)}  \nonumber\\
	&=
	-T \alpha_s t^{-s}\frac{\sin\left(\frac{\pi s}{2}\right)}{\pi^2}\int_0^\infty d\varepsilon\; -2\pi i \left[  
	\frac{ e^{-i(\varepsilon-i0)} (\varepsilon-i0)\varepsilon^{s-1}  }{\varepsilon-i0+i0 +\varepsilon}
	+
	\frac{ e^{-i(-\varepsilon-i0)} (-\varepsilon-i0)\varepsilon^{s-1}  }{-\varepsilon-i0 +i0 -\varepsilon}\right] \nonumber\\
	&=
	2iT \alpha_s t^{-s}\frac{\sin\left(\frac{\pi s}{2}\right)}{\pi}\int_0^\infty d\varepsilon\;  \left[  
	\frac{ e^{-i\varepsilon} \varepsilon^{s}  }{2\varepsilon}
	+
	\frac{ e^{i\varepsilon} \varepsilon^{s}  }{2\varepsilon}\right] \nonumber\\
	&=
	2iT \alpha_s t^{-s}\frac{\sin\left(\frac{\pi s}{2}\right)}{\pi}\int_0^\infty d\varepsilon\;  \cos(\varepsilon) \varepsilon^{s-1}\nonumber\\
	&=
	2iT \alpha_s t^{-s}\frac{\sin\left(\frac{\pi s}{2}\right)}{\pi}\cos\left(\frac{\pi s}{2}\right) \Gamma(s)\nonumber\\
	&=
	iT \alpha_s t^{-s}\frac{\sin\left(\pi s\right)}{\pi} \Gamma(s)\nonumber\\
	&=
	iT \alpha_s \frac{t^{-s}}{\Gamma(1-s)}.
\end{align}
We therefore find that 
\begin{equation}
	\langle \bm{\xi}_m(t) \bm{\xi}_n(t') \rangle = -2i \delta_{m,n}\alpha^K(t-t') = 2\alpha_s  T \delta_{m,n}\frac{ (t-t')^{-s}}{\Gamma(1-s)},	
\end{equation}
where we assumed that $t\ge t'$. Therefore, we have now found the fractional LLG equation
\begin{equation}
	\dot{\bm{S}}(t) = \bm{S}(t) \times \left[ -\bm{B} +\alpha_s D_t^s \bm{S}(t) +\bm{\xi}(t) \right] .
\end{equation}

\subsection{Super-Ohmic spectral function}
We now consider the case where 
\begin{equation}
	J(\varepsilon)=\alpha_s\sin\left(\frac{\pi s}{2}\right) \varepsilon^s,
\end{equation} with $1<s<2$. In this case, everything is equivalent to the sub-Ohmic case, up to  Eq.~\eqref{eq:subohmicmiddle}, where we wanted to rewrite the dissipation into a fractional derivative. We had to introduce a finite initial time $t_0$, which lead to
\begin{align}
	(\adis * \bm{S})(t) &=
	-2\alpha_s\frac{\sin\left(\frac{\pi s}{2}\right)}{\pi} \int_{-\infty}^{t} dt' 
	\int_0^\infty d\varepsilon \sin[\varepsilon(t-t')] \varepsilon^{s}\bm{S}(t')\nonumber\\ 
	&= 
	-2\alpha_s\frac{\sin\left(\frac{\pi s}{2}\right)}{\pi}  
	\int_0^\infty d\varepsilon \left\{
	\left[\varepsilon^{s-1}\cos[\varepsilon(t-t')] \bm{S}(t')\right]_{t'=t_0}^{t'=t}
	-\int_{t_0}^{t} dt' \cos[\varepsilon(t-t')] \varepsilon^{s-1}\dot{\bm{S}}(t')\right\}\nonumber\\ 
	&= 
	-2\alpha_s\frac{\sin\left(\frac{\pi s}{2}\right)}{\pi}  
	\int_0^\infty d\varepsilon \left\{
	\varepsilon^{s-1}\bm{S}(t) -\varepsilon^{s-1}\cos[\varepsilon (t-t_0)] \bm{S}(t_0)
	-\int_{t_0}^{t} dt' \cos[\varepsilon(t-t')] \varepsilon^{s-1}\dot{\bm{S}}(t')\right\}.
\end{align}
The first term vanishes because of the cross product with $\bm{S}(t)$ in the LLG equation. However, the second term is more problematic compared to the sub-Ohmic case, since the identity used to rewrite it only holds for $s<1$. We solve this by writing it as a time derivative
\begin{equation}
	\varepsilon^{s-1}\cos[\varepsilon (t-t_0)] \bm{S}(t_0)= \frac{d}{dt} \varepsilon^{s-2}\sin[\varepsilon (t-t_0)] \bm{S}(t_0),
\end{equation}
and then switching the ordering of the derivative and integral. Performing also one more partial integration in $t'$, we get
\begin{align}
	&(\adis * \bm{S})(t) \nonumber\\
	&=
	2\alpha_s\frac{\sin\left(\frac{\pi s}{2}\right)}{\pi} 
	\int_0^\infty d\varepsilon \left\{\frac{d}{dt} \varepsilon^{s-2}\sin[\varepsilon (t-t_0)] \bm{S}(t_0)
	+\int_{t_0}^t dt' \cos[\varepsilon(t-t')] \varepsilon^{s-1}\dot{\bm{S}}(t')\right\} \nonumber\\
	&=
	2\alpha_s\frac{\sin\left(\frac{\pi s}{2}\right)}{\pi}  
	\int_0^\infty d\varepsilon \left(\frac{d}{dt} \varepsilon^{s-2}\sin[\varepsilon (t-t_0)] \bm{S}(t_0)+	\left[\varepsilon^{s-2}\sin[\varepsilon(t-t')] \dot{\bm{S}}(t')\right]_{t'=t_0}^{t'=t}
	-\int_{t_0}^t dt' \left\{-\sin[\varepsilon(t-t')] \varepsilon^{s-2}\ddot{\bm{S}}(t')\right\}\right) \nonumber\\
	&=
	2\alpha_s\frac{\sin\left(\frac{\pi s}{2}\right)}{\pi}  
	\int_0^\infty d\varepsilon \left\{\frac{d}{dt} \varepsilon^{s-2}\sin[\varepsilon (t-t_0)] \bm{S}(t_0)+	\varepsilon^{s-2}\sin[\varepsilon (t-t_0)] \dot{\bm{S}}(t_0)
	+\int_{t_0}^t dt' \sin[\varepsilon(t-t')] \varepsilon^{s-2}\ddot{\bm{S}}(t')\right\}.
\end{align}
Now, rescaling $\varepsilon\to \varepsilon/(t-t_0)$ and $\varepsilon\to \varepsilon/(t-t')$ respectively, we have
\begin{align}
	&(\adis * \bm{S})(t) \nonumber\\
	&= 
	2\alpha_s\frac{\sin\left(\frac{\pi s}{2}\right)}{\pi}  
	\int_0^\infty d\varepsilon \sin(\varepsilon) \varepsilon^{s-2}\left[ \frac{d}{dt} (t-t_0)^{1-s}\bm{S}(t_0) + (t-t_0)^{1-s}\dot{\bm{S}}(t_0)
	+\int_{t_0}^t dt' (t-t')^{1-s} \ddot{\bm{S}}(t')\right]\nonumber \\
	&= 
	2\alpha_s\frac{\sin\left(\frac{\pi s}{2}\right)}{\pi}  
	\left[\sin\left(\frac{\pi (s-1)}{2}\right) \Gamma(s-1) \right] \left[(1-s)(t-t_0)^{-s}\bm{S}(t_0)+ (t-t_0)^{1-s}\dot{\bm{S}}(t_0)
	+\int_{t_0}^t dt' (t-t')^{1-s} \ddot{\bm{S}}(t')\right]\nonumber \\
	&= 
	-2\alpha_s\frac{\sin\left(\frac{\pi s}{2}\right)}{\pi}  
	\cos\left(\frac{\pi s}{2}\right) \Gamma(s-1)  \left[(1-s)(t-t_0)^{-s}\bm{S}(t_0)+(t-t_0)^{1-s}\dot{\bm{S}}(t_0)
	+\Gamma(2-s) \prescript{C}{t_0}{D}_t^s \bm{S}(t)\right]\nonumber \\
	&= 
	-2\alpha_s\frac{\sin\left(\frac{\pi s}{2}\right)}{\pi}  
	\cos\left(\frac{\pi s}{2}\right) \Gamma(s-1) \Gamma(1-(s-1)) \left[\frac{(t-t_0)^{-s}}{\Gamma(1-s)}\bm{S}(t_0)+ \frac{(t-t_0)^{1-s}}{\Gamma(2-s)}\dot{\bm{S}}(t_0)
	+\prescript{C}{t_0}{D}_t^s \bm{S}(t)\right]\nonumber \\
	&= 
	-\alpha_s\frac{\sin\left(\pi s\right)}{\pi}  
	\frac{\pi}{\sin[\pi (s-1)]} \prescript{RL}{t_0}{D}_t^s \bm{S}(t)\nonumber \\
	&= 
	\alpha_s D_t^s \bm{S}(t),
\end{align}
where we used several identities from Sec. 6 in the Sup. Mat. of Ref.~\cite{Verstraten2021}, Ref.~\cite[p.893]{gradshteyn2014table}, and in the last line we sent $t_0\to-\infty$.

For the noise correlation, we have to compute the Keldysh component. This is
\begin{align}
	\alpha^K(t) &= \frac{1}{2\pi}\int_{-\infty}^\infty d\omega\; e^{-i\omega t} \alpha^K(\omega) \nonumber\\
	&=
	\frac{1}{2\pi}\int_{-\infty}^\infty d\omega\; e^{-i\omega t}[\alpha^R(\omega) -\alpha^A(\omega)]\coth\left( \frac{\omega}{2T} \right)\nonumber\\
	&=
	\alpha_s\frac{\sin\left(\frac{\pi s}{2}\right)}{2\pi^2}\int_0^\infty d\varepsilon\int_{-\infty}^\infty d\omega\; e^{-i\omega t} \left[\frac{ \omega^2\varepsilon^{s-1}  }{\varepsilon^2 -(\omega + i0)^2} -\frac{ \omega^2\varepsilon^{s-1}  }{\varepsilon^2 -(\omega - i0)^2}\right]\coth\left( \frac{\omega}{2T} \right).
\end{align}
Now, we send $\omega\to -\omega$ in the advanced part
	\begin{align}
\alpha^K(t)	&=
	\alpha_s\frac{\sin\left(\frac{\pi s}{2}\right)}{2\pi^2}\int_0^\infty d\varepsilon\int_{-\infty}^\infty d\omega\;  \left[\frac{e^{-i\omega t} \omega^2\varepsilon^{s-1}  }{\varepsilon^2 -(\omega + i0)^2} +\frac{e^{i\omega t} \omega^2\varepsilon^{s-1}  }{\varepsilon^2 -(-\omega - i0)^2}\right]\coth\left( \frac{\omega}{2T} \right)\nonumber\\
	&=
	\alpha_s\frac{\sin\left(\frac{\pi s}{2}\right)}{\pi^2}\int_0^\infty d\varepsilon\int_{-\infty}^\infty d\omega\;  \cos(\omega t)\frac{ \omega^2\varepsilon^{s-1}  }{\varepsilon^2 -(\omega + i0)^2}\coth\left( \frac{\omega}{2T} \right).
\end{align}
Then, we insert $\cos(\omega t) = \frac{d}{dt} \frac{\sin(\omega t)}{\omega}$ and we send $\omega \to \omega/t$ and $\varepsilon\to \varepsilon/t$, which yields
\begin{align}
	\alpha^K(t) &=
	\frac{d}{dt}\alpha_s\frac{\sin\left(\frac{\pi s}{2}\right)}{\pi^2}\int_0^\infty d\varepsilon\int_{-\infty}^\infty d\omega\;  \sin(\omega t)\frac{ \omega \varepsilon^{s-1}  }{\varepsilon^2 -(\omega + i0)^2}\coth\left( \frac{\omega}{2T} \right)\nonumber\\
	&=
	\frac{d}{dt}\alpha_s t^{-s}\frac{\sin\left(\frac{\pi s}{2}\right)}{\pi^2}\int_0^\infty d\varepsilon\int_{-\infty}^\infty d\omega\;  \sin(\omega )\frac{ \omega\varepsilon^{s-1}  }{\varepsilon^2 -(\omega + i0)^2}\coth\left( \frac{\omega}{2tT} \right).
\end{align}
Taking the high-temperature limit, we get
\begin{align}
	\alpha^K(t) &=
	\frac{d}{dt}\alpha_s t^{-s}\frac{\sin\left(\frac{\pi s}{2}\right)}{\pi^2}\int_0^\infty d\varepsilon\int_{-\infty}^\infty d\omega\;  \sin(\omega )\frac{ \omega\varepsilon^{s-1}  }{\varepsilon^2 -(\omega + i0)^2} \frac{2tT}{\omega} \nonumber\\
	&=
	\frac{d}{dt}2T \alpha_s t^{1-s}\frac{\sin\left(\frac{\pi s}{2}\right)}{\pi^2}\int_0^\infty d\varepsilon\int_{-\infty}^\infty d\omega\;  \sin(\omega )\frac{ \varepsilon^{s-1}  }{\varepsilon^2 -(\omega + i0)^2} \nonumber\\
	&=
	-2T \alpha_s (1-s)t^{-s}\frac{\sin\left(\frac{\pi s}{2}\right)}{\pi^2}\int_0^\infty d\varepsilon\int_{-\infty}^\infty d\omega\;  \frac{ \sin(\omega )\varepsilon^{s-1}  }{(\omega +i0 -\varepsilon)(\omega+i0 +\varepsilon)}.  
\end{align}
Now, we want to close the integral over $\omega$ with an infinite half-circle. For this, we need fast enough convergence of the integrand to zero. Splitting the cosine into two exponential parts $\sin(\omega)= (e^{i\omega} -e^{-i\omega})/2i$, we see that the first term goes to zero when $\omega\to i\infty$ and that the second term goes to zero when $\omega\to-i\infty$. Since we have poles at $\omega=\pm\varepsilon -i0$, the integral along the top half-plane vanishes. The integral along the bottom is then computed as
\begin{align}
	\alpha^K(t) &=
	-i T \alpha_s (1-s)t^{-s}\frac{\sin\left(\frac{\pi s}{2}\right)}{\pi^2}\int_0^\infty d\varepsilon\int_{-\infty}^\infty d\omega\;  \frac{ e^{-i\omega} \varepsilon^{s-1}  }{(\omega +i0 -\varepsilon)(\omega+i0 +\varepsilon)}  \nonumber\\
	&=
	-i T \alpha_s (1-s)t^{-s}\frac{\sin\left(\frac{\pi s}{2}\right)}{\pi^2}\int_0^\infty d\varepsilon\; -2\pi i \left[  
	\frac{ e^{-i(\varepsilon-i0)} \varepsilon^{s-1}  }{\varepsilon-i0+i0 +\varepsilon}
	+
	\frac{ e^{-i(-\varepsilon-i0)} \varepsilon^{s-1}  }{-\varepsilon-i0 +i0 -\varepsilon}\right] \nonumber\\
	&=
	-2T \alpha_s (1-s) t^{-s}\frac{\sin\left(\frac{\pi s}{2}\right)}{\pi}\int_0^\infty d\varepsilon\;  \left[  
	\frac{ e^{-i\varepsilon} \varepsilon^{s-2}  }{2}
	-
	\frac{ e^{i\varepsilon} \varepsilon^{s-2}  }{2}\right] \nonumber\\
	&=
	i 2T \alpha_s (1-s) t^{-s}\frac{\sin\left(\frac{\pi s}{2}\right)}{\pi}\int_0^\infty d\varepsilon\;  \sin(\varepsilon) \varepsilon^{s-2}\nonumber\\
	&=
	-i 2T \alpha_s (1-s) t^{-s}\frac{\sin\left(\frac{\pi s}{2}\right)}{\pi}\cos\left(\frac{\pi s}{2}\right) \Gamma(s-1)\nonumber\\
	&=
	i T \alpha_s  t^{-s}\frac{\sin\left(\pi s\right)}{\pi} \Gamma(s)\nonumber\\
	&=
	iT \alpha_s \frac{t^{-s}}{\Gamma(1-s)}.
\end{align}
We therefore find the same expression as in the sub-Ohmic case, which leads to
\begin{equation}
	\langle \bm{\xi}_m(t) \bm{\xi}_n(t') \rangle = -2i \delta_{m,n}\alpha^K(t-t') = 2\alpha_s T \delta_{m,n}\frac{ (t-t')^{-s}}{\Gamma(1-s)}.	
\end{equation}
Therefore, we have now also found the fractional LLG equation in the super-Ohmic case.

\subsection{Comparison Ohmic versus non-Ohmic}
In the Ohmic case, we started with $J(\omega)= \alpha_1\omega$ and ended up with a $\alpha_1 \dot{\bm{S}}(t)$ friction term, and noise correlation $4\alpha_1 k_B T \delta(t-t')$. On the other hand, in the non-Ohmic case, we started with $J(\varepsilon)=\alpha_s\sin\left(\frac{\pi s}{2}\right) \varepsilon^s$ and ended up with a friction $\alpha_s D_{t}^s \bm{S}(t)$ and noise correlation $2\alpha_s k_B T \frac{ (t-t')^{-s}}{\Gamma(1-s)}$. Although it is clear that both the non-Ohmic $J(\omega)$ and the friction term will go to the Ohmic case when $s\to 1$, the noise is less straightforward. We can see that, as $s\to 1$, the Gamma function will blow up, hence sending the correlation to zero, with the exception of $t=t'$. In this case, the numerator blows up even before taking the limit of $s \to 1$. Hence, we can expect this function to behave as a delta function. To find the correct prefactor, we integrate the distribution with a test-function $f(t')=1$ to find 
\begin{align*}
	\lim_{s\to 1}\int_{-\infty}^{\infty} d(t-t') 	\frac{ (t-t')^{-s}}{\Gamma(1-s)}\cdot 1 &=
	\lim_{s\to 1}	\left[\frac{ (t-t')^{1-s}}{(1-s)\Gamma(1-s)}\right]_{(t-t')=-\infty}^{\infty}\\
	&=
	\lim_{s\to 1}	\left[\frac{ (t-t')^{1-s}}{\Gamma(2-s)}\right]_{(t-t')=-\infty}^{\infty}\\
	&= 
	\left[\frac{ (t-t')}{|t-t'|\Gamma(1)}\right]_{(t-t')=-\infty}^{\infty}\\
	&=2.
\end{align*}
Hence, we see that the limit of the noise correlation becomes $ 4\alpha_1 k_B T \delta(t-t')$, which is precisely as in the Ohmic case.

\section{FMR powerlaw derivation}

Here, we calculate the response of this spin-bath system to a rotating magnetic field. We will find the steady state solutions and calculate several quantities that experiments could measure. 

\subsection{Ferromagnetic Resonance}
We will study the effects of a rotating magnetic field on the fractional LLG (FLLG) equation
\begin{equation}
	\dot{\bm{S}}(t) = \bm{S}(t) \times \big\{ -\bm{B}_{\text{eff}}[t,\bm{S}(t)] + \alpha_s D_t^s \bm{S}(t) +\bm{\xi}(t) \big\}.
\end{equation}
For simplicity, we assume that the temperature of the bath is low compared to the energy of the external fields, such that the thermal noise $\bm{\xi}(t)$ may be neglected. We apply a rotating magnetic field
\begin{equation}
	\bm{B}_{\text{eff}}[t,\bm{S}(t)]= \begin{pmatrix}
		\Omega \cos(\omega_d t)\\ \Omega \sin(\omega_d t)\\B_0-K S_z(t)
	\end{pmatrix}
\end{equation}
and use spherical coordinates 
\begin{equation}
	\bm{S} = S \begin{pmatrix}
		\sin\theta \cos\phi \\ \sin\theta \sin\phi \\ \cos\theta
	\end{pmatrix}.
\end{equation}
 We will assume a small $\theta$ approximation, where the ground state is in the positive $z$ direction, i.e.  $0<\Omega \ll B_0-K S$ and $\alpha_s S \ll (B_0-K S)^{1-s}$. As shown in Section~\ref{sec:sphericalllg}, we may rewrite the FLLG equation of this form in spherical coordinates as
\begin{align}
	\dot{\phi} &=  \frac{1}{\sin\theta}
	\left[ -\bm{B}_{\text{eff}}[t,\bm{S}(t)] + \alpha_s D_t^s \bm{S}(t)	\right]\cdot  \begin{pmatrix}
		\cos\theta \cos\phi \\ \cos\theta \sin\phi \\ -\sin\theta
	\end{pmatrix}; 
	\\
	\dot{\theta}   &= 
	\left[ -\bm{B}_{\text{eff}}[t,\bm{S}(t)] + \alpha_s D_t^s \bm{S}(t)	\right]\cdot  \begin{pmatrix}
		\sin\phi \\ -\cos\phi \\ 0
	\end{pmatrix}. 
\end{align}
In the rotating frame, where $\bm{B}(t)$ is constant, we could expect the system to go to a steady state after some time. Hence, we introduce a new coordinate such, that $\phi= \omega_d t-\varphi$. We may then set $\dot{\varphi}=\dot{\theta}=0$ to find the steady state in the rotating frame, where
\begin{align}
	\omega_d \sin\theta &=  
	\left[ -\bm{B}_{\text{eff}}[t,\bm{S}(t)] + \alpha_s D_t^s \bm{S}(t)	\right]\cdot  \begin{pmatrix}
		\cos\theta \cos(\omega_d t-\varphi) \\ \cos\theta \sin(\omega_d t-\varphi) \\ -\sin \theta
	\end{pmatrix}; 
	\\
	0   &= 
	\left[ -\bm{B}_{\text{eff}}[t,\bm{S}(t)] + \alpha_s D_t^s \bm{S}(t) 	\right]\cdot  \begin{pmatrix}
		\sin(\omega_d t-\varphi) \\ -\cos(\omega_d t-\varphi) \\ 0
	\end{pmatrix}. 
\end{align}
We note that $\bm{S}(t)$ is now only time dependent in the rotating-frame term, which means that we can explicitly calculate the fractional derivative. The Liouville derivative works well with Fourier transforms, hence the fractional derivative of a trigonometric function is given by 
\begin{equation}
	D_t^s \sin(\omega t)= |\omega|^s \sin\left( \omega t +\text{sign}(\omega)\frac{\pi s}{2}\right),
\end{equation}
and similarly for a cosine. We remark that the Liouville derivative of a constant can only be described by setting the initial time to some finite $t_0$, in which case it becomes zero\footnote{Since physically the infinite time only models a long time in the past, we will apply it as such. Formally, there are some restrictions on the functions that the Liouville derivative can be applied to. These include restrictions such as its integral over the whole domain being finite. Since this is not the case for a non-zero constant on an infinite interval, we have to regularize the lower integral boundary $-\infty$ as a finite $t_0$.}. Combining this with the steady state expression for $\bm{S}$, we find
\begin{equation}
	\alpha_s D_t^s \bm{S}(t)= \alpha_s D_t^s S \begin{pmatrix}
		\sin\theta \cos(\omega_d t -\varphi) \\\sin\theta \sin(\omega_d t -\varphi)\\ \cos\theta
	\end{pmatrix}
	=
	\alpha_s  S\sin\theta |\omega_d|^s \begin{pmatrix}
		\cos\left(\omega_d t -\varphi+\text{sign}(\omega_d)\frac{\pi s}{2}\right) \\ \sin\left(\omega_d t -\varphi+\text{sign}(\omega_d)\frac{\pi s}{2}\right)\\ 0
	\end{pmatrix}.
\end{equation}
Hence, we find that
\begin{align}
	\omega_d\tan\theta  &=  
	\left[ -\bm{B}_{\text{eff}}[t,\bm{S}(t)] + \alpha_s D_t^s \bm{S}(t)	\right]\cdot  \begin{pmatrix}
		\cos(\omega_d t -\varphi) \\  \sin(\omega_d t -\varphi) \\ -\tan\theta
	\end{pmatrix} \nonumber
	\\
	&=
	\left[-\begin{pmatrix}
		\Omega \cos(\omega_d t)\\ \Omega \sin(\omega_d t)\\B_0 -K S \cos\theta
	\end{pmatrix} +\alpha_s  S\sin\theta |\omega_d|^s \begin{pmatrix}
		\cos\left(\omega_d t -\varphi+\text{sign}(\omega_d)\frac{\pi s}{2}\right) \\ \sin\left(\omega_d t -\varphi+\text{sign}(\omega_d)\frac{\pi s}{2}\right)\\ 0
	\end{pmatrix}\right]\cdot  \begin{pmatrix}
		\cos(\omega_d t -\varphi) \\  \sin(\omega_d t -\varphi) \\ -\tan\theta
	\end{pmatrix}\nonumber \\
	&=
	-\Omega\cos\varphi +B_0\tan\theta -K S \sin\theta+\alpha_s  S \sin\theta|\omega_d|^s\cos\left(\text{sign}(\omega_d)\frac{\pi s}{2}\right)
\end{align}
and 
\begin{align}
	0   &= 
	\left[-\bm{B}_{\text{eff}}[t,\bm{S}(t)] + \alpha_s D_t^s \bm{S}(t) 	\right]\cdot  \begin{pmatrix}
		\sin(\omega_d t -\varphi) \\ -\cos(\omega_d t -\varphi) \\ 0
	\end{pmatrix}\nonumber\\
	&=
	\left[-\begin{pmatrix}
		\Omega \cos(\omega_d t)\\ \Omega \sin(\omega_d t)\\B_0 -K S \cos\theta
	\end{pmatrix} 
	+\alpha_s  S\sin\theta |\omega_d|^s \begin{pmatrix}
		\cos\left(\omega_d t -\varphi+\text{sign}(\omega_d)\frac{\pi s}{2}\right) \\ \sin\left(\omega_d t -\varphi+\text{sign}(\omega_d)\frac{\pi s}{2}\right)\\ 0
	\end{pmatrix}\right]\cdot  \begin{pmatrix}
		\sin(\omega_d t -\varphi) \\ -\cos(\omega_d t -\varphi) \\ 0
	\end{pmatrix}\nonumber\\
	&=
	\Omega \sin\varphi -\alpha_s  S\sin\theta |\omega_d|^s \sin\left(\text{sign}(\omega_d)\frac{\pi s}{2}\right).
\end{align} 
With some rearranging, we have 
\begin{equation}
	\varphi= \arcsin\left[\frac{\alpha_s S}{\Omega}   |\omega_d|^s \sin\left(\text{sign}(\omega_d)\frac{\pi s}{2}\right)\sin\theta\right]
\end{equation}
and 
\begin{align}
\Omega\cos\varphi
&=
 \Omega \cos\arcsin\left[\frac{\alpha_s S}{\Omega}   |\omega_d|^s \sin\left(\text{sign}(\omega_d)\frac{\pi s}{2}\right)\sin\theta\right]\nonumber\\	
&=
\sqrt{\Omega^2- \left[\alpha_s S|\omega_d|^s \sin\left(\text{sign}(\omega_d)\frac{\pi s}{2}\right)\sin\theta\right]^2}  \nonumber\\
&=	
(B_0-K S \cos\theta-\omega_d) \tan\theta  +\alpha_s  S \sin\theta|\omega_d|^s\cos\left(\text{sign}(\omega_d)\frac{\pi s}{2}\right).
\end{align}
Squaring this, we get
\begin{align*}
	\Omega^2	  &=(B_0-K S \cos\theta-\omega_d)^2 \tan^2\theta +\left(\alpha_s  S |\omega_d|^s\right)^2\sin^2\theta +2\alpha_s  S  |\omega_d|^s(B_0-K S \cos\theta-\omega_d)\cos\left(\frac{\pi s}{2}\right)\frac{\sin^2\theta}{\cos\theta},
\end{align*}
and multiplying by $\cos^2\theta=1-\sin^2\theta$, we have
\begin{align}
	(1-\sin^2\theta)\Omega^2&=
	(B_0-K S \sqrt{1-\sin^2\theta}-\omega_d)^2 \sin^2\theta +\left[\left(\alpha_s  S |\omega_d|^s\right)^2-2\alpha_s  S  |\omega_d|^s K S \cos\left(\frac{\pi s}{2}\right)\right]\sin^2\theta(1-\sin^2\theta) \nonumber\\
	&\quad+2\alpha_s  S  |\omega_d|^s(B_0-\omega_d)\cos\left(\frac{\pi s}{2}\right)\sin^2\theta \sqrt{1-\sin^2\theta}.
\end{align}
We could go further and make this into (effectively) a 4th order equation for $\sin^2\theta$. However, since we are in a small $\theta$ limit, we will solve this equation up to first order in $\sin^2\theta$, which yields
\begin{align}
	\Omega^2&=\sin^2\theta\Big\{ \Omega^2
	(B_0-K S -\omega_d)^2  +\left[\left(\alpha_s  S |\omega_d|^s\right)^2-2\alpha_s  S  |\omega_d|^s K S \cos\left(\frac{\pi s}{2}\right)\right]  \nonumber\\
	&\quad+2\alpha_s  S  |\omega_d|^s (B_0-\omega_d)\cos\left(\frac{\pi s}{2}\right) \Big\}.
\end{align}
Hence, we find that
\begin{align}
\sin^2\theta &= 
 \frac{\Omega^2}{\Omega^2 +
	(B_0-K S -\omega_d)^2  +\left(\alpha_s  S |\omega_d|^s\right)^2 +2\alpha_s  S  |\omega_d|^s (B_0-K S-\omega_d)\cos\left(\frac{\pi s}{2}\right)}\nonumber\\
&=
\frac{\Omega^2}{
	(B_0-K S -\omega_d)^2  +\left(\alpha_s  S |\omega_d|^s\right)^2 +2\alpha_s  S  |\omega_d|^s (B_0-K S-\omega_d)\cos\left(\frac{\pi s}{2}\right)} +\mathcal{O}(\Omega^4). \label{eq:simplesin}
\end{align}
Note that the $\mathcal{O}(\Omega^4)$ should formally be dimensionless, but we explain what we mean with terms being small in Section~\ref{sec:dim}, as this is more subtle with fractional dimensions.

\subsection{Resonance frequency and amplitude}
Since we are studying ferromagnetic resonance, we want to find the driving frequency for which we get the largest response from the magnetic system. Since we know that the resonance for an Ohmic system is at $\omega_d=B_0-K S$, we will expand the formula around this point to find the new maximum. To compute the resonance frequency $\omega_\text{res}$, we thus first assume that $\omega_\text{res} \approx (B_0-KS)(1+y)$ with $y$ small, such that $|\omega_\text{res}|^s \approx (B_0-KS)^s (1+ sy)$ (for $B_0-K S>0$). This results in 
\begin{align}
	\sin^2\theta 
	&=
	\frac{\Omega^2}{
		(B_0-K S -\omega_d)^2  +\left(\alpha_s  S |\omega_d|^s\right)^2 +2\alpha_s  S  |\omega_d|^s (B_0-K S-\omega_d)\cos\left(\frac{\pi s}{2}\right)} \nonumber\\
	&\approx
	\frac{\Omega^2}{
		(B_0-K S)^2 y^2  +\left(\alpha_s  S \right)^2 (B_0-KS)^{2s}(1+2s y) -2\alpha_s  S  (B_0-KS)^{s+1} (1+s y)y \cos\left(\frac{\pi s}{2}\right)}.
\end{align}
Now, we put the derivative with respect to $y$ equal to zero, to get
\begin{equation}
	(B_0-K S)^2 y + s\left(\alpha_s  S \right)^2 (B_0-KS)^{2s}  -\alpha_s  S  (B_0-KS)^{s+1} (1+2s y) \cos\left(\frac{\pi s}{2}\right) =0.
\end{equation}
Hence, we find that 
\begin{align}
	y
	&=
	\frac{-s\left(\alpha_s  S \right)^2 (B_0-KS)^{2s}+\alpha_s  S  (B_0-KS)^{s+1}  \cos\left(\frac{\pi s}{2}\right)}{(B_0-K S)^2-2s\alpha_s  S  (B_0-KS)^{s+1}  \cos\left(\frac{\pi s}{2}\right)}\nonumber\\
	&=
	\alpha_s  S  (B_0-KS)^{s-1}  \cos\left(\frac{\pi s}{2}\right) +\mathcal{O}(\alpha_s S)^2,
\end{align}
which results in
\begin{equation}
	\omega_\text{res} \approx (B_0-KS)\left[1+\alpha_s  S  (B_0-KS)^{s-1}  \cos\left(\frac{\pi s}{2}\right)\right]
	=(B_0-KS) +\alpha_s  S  (B_0-KS)^{s}  \cos\left(\frac{\pi s}{2}\right).
\end{equation}
We see that the resonance frequency gets shifted by a small amount, depending on $s$, which scales non-linearly. 
Inserting this result into Eq. \eqref{eq:simplesin}, we can now also find an approximation for the amplitude at resonance:
\begin{align}
\sin^2\theta_\text{res} 
&=
\frac{\Omega^2}{
	(B_0-K S -\omega_\text{res} )^2  +\left(\alpha_s  S |\omega_\text{res} |^s\right)^2 +2\alpha_s  S  |\omega_\text{res} |^s (B_0-K S-\omega_\text{res} )\cos\left(\frac{\pi s}{2}\right)} \nonumber\\
&\approx
\Omega^2\Big\{
	\left[\alpha_s  S  (B_0-KS)^{s}  \cos\left(\frac{\pi s}{2}\right) \right]^2  +\left(\alpha_s  S \left|(B_0-KS) +\alpha_s  S  (B_0-KS)^{s}  \cos\left(\frac{\pi s}{2}\right) \right|^s\right)^2 \nonumber\\
	&\quad-2\alpha_s  S  \left|(B_0-KS) +\alpha_s  S  (B_0-KS)^{s}  \cos\left(\frac{\pi s}{2}\right) \right|^s \alpha_s  S  (B_0-KS)^{s}  \cos^2\left(\frac{\pi s}{2}\right)\Big\}^{-1} \nonumber\\
	&=
	\frac{\Omega^2}{\left[\alpha_s S(B_0-KS)^{s}\right]^2}\Big[
	 \cos^2\left(\frac{\pi s}{2}\right)   +\left|1 +\alpha_s  S  (B_0-KS)^{s-1}  \cos\left(\frac{\pi s}{2}\right) \right|^{2s} \nonumber\\
	&\quad-2 \left|1 +\alpha_s  S  (B_0-KS)^{s-1}  \cos\left(\frac{\pi s}{2}\right) \right|^s    \cos^2\left(\frac{\pi s}{2}\right) \Big]^{-1} \nonumber\\
		&=
	\frac{\Omega^2}{\left[\alpha_s S(B_0-KS)^{s}\right]^2}\Big[1-
	\cos^2\left(\frac{\pi s}{2}\right) +\mathcal{O}(\alpha_s S) \Big]^{-1} \nonumber\\
	&\approx
	\frac{\Omega^2}{\left[\alpha_s S(B_0-KS)^{s}\sin\left(\frac{\pi s}{2}\right)\right]^2}.
\end{align}
Since the sine function decreases as $s$ moves away from one, we see that the amplitude actually increases for non-Ohmic environments.

\subsection{Calculating the FWHM linewidth}
Next, we are interested not only in the location of the resonance, but also how sensitive the resonance is to the driving frequency. One way to describe this is by using the Full Width at Half Maximum measure. This provides a well-defined line width independently of the shape of the peak. It is found by measuring the width of the peak at half the height of its maximum. This can be measured in the laboratories, but it can also be computed. Since our function of interest is of the form $\sin^2\theta(\omega_d) =\Omega^2 /g(\omega_d)$, it makes sense to approximate the inverse function instead of the regular one. To this end, we will translate the FWHM measurement to the inverse function, and then Taylor expand $g(\omega_d)$ near resonance as a parabola to solve for the new condition of this inverse function. Notice that from Eq. \eqref{eq:simplesin}, we have
\begin{equation}
	g(\omega_d)=
	(B_0-K S -\omega_d)^2  +\left(\alpha_s  S |\omega_d|^s\right)^2 +2\alpha_s  S  |\omega_d|^s (B_0-K S-\omega_d)\cos\left(\frac{\pi s}{2}\right).
\end{equation}
The FWHM condition is 
\begin{equation}
	\frac{\Omega^2 }{g(\omega_d)}=	\sin^2\theta(\omega_d) = \frac{\sin^2\theta(\omega_\text{res})}{2} =\frac{\Omega^2 }{2g(\omega_\text{res})},
\end{equation}
hence we must solve for $2g(\omega_\text{res})=g(\omega_d)$. To this end, let us assume that $\omega_d = \omega_\text{res} +y$ and expand $g(\omega_d)$ in $y$. We will use that 
$$|a+y|^n \approx a^n +n a^{n-1} y +\frac{1}{2} n(n-1)a^{n-2} y^2$$
 for small $y$ and $a>0$. Then,
\begin{align}
	&g(\omega_\text{res}+y)	\nonumber\\
	&=
	(B_0-K S -\omega_\text{res}-y)^2  +\left(\alpha_s  S |\omega_\text{res}+y|^s\right)^2 +2\alpha_s  S  |\omega_\text{res}+y|^s (B_0-K S-\omega_\text{res}-y)\cos\left(\frac{\pi s}{2}\right)\nonumber\\
	&\approx
	(B_0-K S -\omega_\text{res})^2  +\left(\alpha_s  S  \omega_\text{res}^s\right)^2 +2\alpha_s  S  \omega_\text{res}^s (B_0-K S-\omega_\text{res})\cos\left(\frac{\pi s}{2}\right) \nonumber\\
	&+y\left( -2(B_0-K S -\omega_\text{res})  +2s(\alpha_s  S)^2 \omega_\text{res}^{2s-1}  -2\alpha_s  S \cos\left(\frac{\pi s}{2}\right)\left\{ \omega_\text{res}^{s} +s\omega_\text{res}^{s-1}[\omega_\text{res}- (B_0-K S)]\right\} \right)\nonumber\\
	&+y^2\left[ 1 +s(2s-1)(\alpha_s  S )^2\omega_\text{res}^{2s-2} 
	-2s\alpha_s  S  \omega_\text{res}^{s-1} \cos\left(\frac{\pi s}{2}\right) 
	+s(s-1)\alpha_s  S   \omega_\text{res}^{s-2}  (B_0-K S-\omega_\text{res})\cos\left(\frac{\pi s}{2}\right) \right]\nonumber\\
	&=
	g(\omega_\text{res}) +y\Bigg( 2\alpha_s  S  (B_0-KS)^{s}  \cos\left(\frac{\pi s}{2}\right)  +2s(\alpha_s  S)^2 (B_0-KS)^{2s-1}\left[1+\alpha_s  S  (B_0-KS)^{s-1}  \cos\left(\frac{\pi s}{2}\right)\right]^{2s-1}\nonumber\\
	&\qquad-2\alpha_s  S (B_0-KS)^{s} \cos\left(\frac{\pi s}{2}\right)\Bigg\{
	\left[1+\alpha_s  S  (B_0-KS)^{s-1}  \cos\left(\frac{\pi s}{2}\right)\right]^{s}\nonumber\\
	&\qquad\qquad+s\alpha_s  S    \cos\left(\frac{\pi s}{2}\right)(B_0-KS)^{s-1}\left[1+\alpha_s  S  (B_0-KS)^{s-1}  \cos\left(\frac{\pi s}{2}\right)\right]^{s-1}
	\Bigg\} \Bigg)\nonumber\\
	&+y^2\Bigg\{ 1 +s(2s-1)(\alpha_s  S )^2(B_0-KS)^{2s-2}\left[1+\alpha_s  S  (B_0-KS)^{s-1}  \cos\left(\frac{\pi s}{2}\right)\right]^{2s-2} \nonumber\\
	&\qquad-2s\alpha_s  S  (B_0-KS)^{s-1}\cos\left(\frac{\pi s}{2}\right) \left[1+\alpha_s  S  (B_0-KS)^{s-1}  \cos\left(\frac{\pi s}{2}\right)\right]^{s-1} \nonumber\\
	&\qquad-s(s-1)(\alpha_s  S)^2   (B_0-KS)^{2s-2} \cos^2\left(\frac{\pi s}{2}\right)\left[1+\alpha_s  S  (B_0-KS)^{s-1}  \cos\left(\frac{\pi s}{2}\right)\right]^{s-2}   \Bigg\}\nonumber\\
	&\approx
	g(\omega_\text{res}) +y\left\{2s(\alpha_s  S)^2 (B_0-KS)^{2s-1}\left[1-2\cos^2\left(\frac{\pi s}{2}\right)\right]\right\} \nonumber\\
	&+y^2\Big\{ 1 -2s(\alpha_s S)(B_0-KS)^{s-1}\cos\left(\frac{\pi s}{2}\right) +(\alpha_s  S )^2(B_0-KS)^{2s-2} \left[ s(2s-1) -3s(s-1)\cos^2\left(\frac{\pi s}{2}\right) \right]
	\Big\} +\mathcal{O}(\alpha_s S)^3.
\end{align}	
Now, we set 
$2g(\omega_\text{res})=g(\omega_\text{res}+y) = g(\omega_\text{res}) +by +a y^2$, and remark that $g(\omega_\text{res})=(\alpha_s S)^2(B_0-KS)^{2s}\sin^2\left(\frac{\pi s}{2}\right)$,
in order to find that 
\begin{align}
	y&= \frac{-b \pm \sqrt{b^2+4a g(\omega_\text{res})}	}{2a} \Longrightarrow \Delta_{FWHM} = \frac{\sqrt{b^2+4a g(\omega_\text{res})}	}{a}.
\end{align}
Hence, we find that the lowest-order contribution to the linewidth is given by
\begin{align}
\Delta_{FWHM}	&\approx \frac{\sqrt{4(\alpha_s S)^2(B_0-KS)^{2s}\sin^2\left(\frac{\pi s}{2}\right)+\mathcal{O}(\alpha_s S)^3}}{1+\mathcal{O}(\alpha_s S)}\nonumber\\
	&= 2(\alpha_s S)(B_0-KS)^{s}\sin\left(\frac{\pi s}{2}\right) + \mathcal{O}(\alpha_s S)^2.
\end{align}

\section{Dimensional analysis}\label{sec:dim}
The fractional derivative in the LLG equation has an impact on the dimensions of quantities. We can firstly see that in the chosen units, we have $[B_0-KS]=[\omega_d]=\text{time}^{-1}$. Assuming $S$ to be dimensionless, then $[\omega_d]=[\alpha_s D_t^s \bm{S}]=[\alpha_s] [\omega_d]^{s}$, hence $[\alpha_s]= [\omega_d]^{1-s}$. We can now start to understand what we mean when we say that certain quantities are small, since this has to be relative to something else. For instance, when we say $\alpha_s S$ is small, we understand this as $\alpha_s S \ll (B_0-KS)^{1-s}$. For $\Omega$ it is simpler, since there is no fractional derivative acting with it. Hence, for $\Omega$ small we simply mean $\Omega\ll B_0-KS$. We can now also define some dimensionless variables, such as $\alpha_s'= \alpha_s S (B_0-KS)^{s-1}$ and $\Omega'= \Omega /(B_0-KS)$. We have used these variables in the figures to show the general behavior of the quantities.

\bibliographystyle{apsrev}
\bibliography{supplementary}

\clearpage
\appendix